\documentclass[11pt,letterpaper]{article}
\usepackage[margin=0.5in]{geometry}
\usepackage{amsmath,amsfonts,amssymb}
\usepackage{pifont}\usepackage{wrapfig}
\usepackage{algorithmic}
\usepackage{algorithm}
\usepackage{array}
\usepackage{textcomp}
\usepackage{stfloats}
\usepackage{url}
\usepackage{verbatim}
\usepackage{graphicx}
\usepackage{multirow}
\usepackage{xcolor}
\usepackage{romannum}
\usepackage{multirow}
\usepackage{bm}
\usepackage{colortbl}
\usepackage{color}
\usepackage{cite}
\definecolor{tablecolor}{rgb}{0.94,0.94,0.94}

\usepackage[labelfont=bf,figurename=Fig]{caption}

\title{Power-Area Efficient Serial IMPLY-based 4:2 Compressor Applied in Data-Intensive Applications\thanks{Preprint Submitted to ArXiV}}

\author{Bahareh Bagheralmoosavi$^{1,}$\thanks{First Author: Bahareh Bagheralmoosavi (b.mousavi@eng.ui.ac.ir)}~, Seyed Erfan Fatemieh$^{1,}$\thanks{Other Author: Seyed Erfan Fatemieh (erfanfatemieh@eng.ui.ac.ir)}~, Mohammad Reza Reshadinezhad$^{1,}$\thanks{Corresponding author: Mohammad Reza Reshadinezhad (m.reshadinezhad@eng.ui.ac.ir)}~, \\and Antonio Rubio$^{2,}$\thanks{Other Author: Antonio Rubio (antonio.rubio@upc.edu)}}

\begin{document}
\maketitle
	
{\noindent$^{1}$ \small \textit{Department of Computer Architecture, Faculty of Computer Engineering, University of Isfahan, Isfahan 8174673441, Iran}}

{\noindent$^{2}$ \small \textit{Department of Electronics Engineering, Universitat Politecnica de Catalunya, Barcelona, Spain}}

\section*{Abstract}
The data transfer between a processor and memory has become a design bottleneck in data-intensive applications. Processing-In-Memory (PIM) is a practical approach to overcome the memory wall bottleneck. The 4:2 compressor is suitable for implementing the processor’s crucial arithmetic circuits, including multiplier. Some area-efficient memristive structures, like Material Implication (IMPLY) in serial architecture, are compatible with the crossbar array. This paper proposes a serial memristive IMPLY-based 4:2 compressor, which is applied to present new 4-bit and 8-bit multipliers. The proposed circuits are evaluated regarding latency, area, and energy consumption. Compared to the existing serial compressor, the proposed 4:2 compressor’s algorithm improves the area, energy consumption, and speed by 36\%, 17\%, and 15\%, respectively. The proposed 4-bit and 8-bit multipliers are improved by 7.3\% and 10\%, respectively, regarding the latency, and reduced energy consumption by up to 12\%, compared to the serial multiplier based on a 4:2 compressor with XOR/MUX design.

\subsection*{Keywords}
In-Memory Computing, Processing-In-Memory, IMPLY logic, 4:2 Compressor, Multiplier, Memristor.

\section{Introduction} \label{sec1}
Today, industry and academia face complex and large-scale processing applications. Processors play an essential role in meeting the requirements of this field. The memory wall has become a challenge in designing efficient processing systems \cite{ref1}. In the Von-Neumann architecture, data is processed and moves between the main memory and the Central Processing Unit (CPU). The speed difference between the CPU and the main memory causes a memory wall bottleneck due to the slower speed of the main memory \cite{ref43,rref46}. One existing approach to the memory wall problem is PIM. In this way, the overhead of data passing between the CPU and the main memory is eliminated, reducing processing time and energy consumption \cite{ref1, ref2}. One of the technologies applied to implement arithmetic circuits is the memristor, which has received much attention for PIM as well \cite{rref47}. Emerging memories, including Resistive Random Access Memory (RRAM), Spin Transfer Torque Magnetic RAM (STT-MRAM), and Phase Change Memory (PCM), are memristive devices with varying properties relating to write time, energy consumption, and endurance that make them suitable for PIM \cite{rref45, rref48, rref49}. Companies such as Fujitsu and Panasonic have commercialized metal-oxide memristors \cite{rref45}. These memristors are constructed by sandwiching metal-oxide (such as $TiO_{2}$ and $HfO_{2}$ with two different doped regions) between two metal electrodes, like platinum, as the conductive terminals \cite{rref49}. These devices' resistance changes are based on the oxygens ions' deficiency or excess \cite{rref49}. The length of the conductive filament determines the memristor's High Resistance State ($R_{off}$) and the Low Resistance State ($R_{on}$) \cite{rref49}. These non-volatile memory devices can store data while performing logical operations. Other advantages of the memristor are high density and low area occupation \cite{ref3}. Basic logic gates and memristor-based arithmetic circuits can be implemented by applying different design methods, including IMPLY \cite{ref4}, MAGIC (Memristor Aided loGIC) \cite{ref5}, FELIX (Fast and Energy-Efficient Logic) \cite{ref6}, and MRL (Memristor Ratioed Logic) \cite{ref7}. IMPLY logic is a design method that is entirely based on memristor devices. It is also a stateful logic, which means the memristor's resistance determines the circuit's input and output logic states. In IMPLY logic, there are no read/write operations to execute logical operations and computations \cite{ref4, ref8}. Some implementation architectures of IMPLY logic, including the serial one, are compatible with the crossbar array structure. Thus, it applies to PIM \cite{ref9}.

\begin{figure}[h]
	\centering
	\includegraphics[scale=0.4]{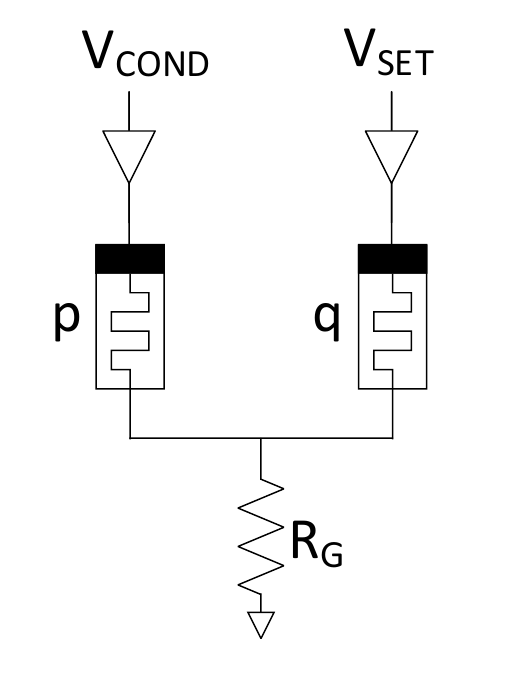}
	\caption{Memristor-based IMPLY gate \cite{ref28}.}
	\label{fig1}
\end{figure}

Fig. \ref{fig1} shows a circuit diagram of the IMPLY circuit and its truth table tabulated in Table \ref{tab1}, followed by its logic function written in (\ref{eq1}). IMPLY logic circuit consists of one $R_{G}$ resistor ($R_{ON}<R_{G}<R_{OFF}$), coupling with two memristors, $p$ and $q$ \cite{ref4, arxiv1}. $V_{COND}$ and $V_{SET}$ voltages are conducted to memristors $p$ and $q$, respectively. Applying condition $|V_{COND}|<|V_{SET}|$ is essential for the correct switching of memristors and the proper function of the IMPLY logic \cite{ref4, arxiv1}. The logic value of the gate inputs determines the initial logic state of memristors $p$ and $q$. At the end of the operation, the final logic state of memristor $q$ equals the logic value of the output \cite{ref4}. More details about the IMPLY logic design method are available in \cite{ref4, ref26}.

\begin{equation}
    p \to q \equiv \overline{p}+q
    \label{eq1}
\end{equation}

\begin{table}[h]
	\centering
	\caption{Truth table of IMPLY function \cite{ref4}}
	\scalebox{1}{
		\begin{tabular}{|c|c|c|c|}
			\hline
			\textbf{Case} & $p$ & $q$ & $p\to q$ \\ \hline
			1 & 0 & 0 & 1 \\ \hline
			2 & 0 & 1 & 1 \\ \hline
			3 & 1 & 0 & 0 \\ \hline
			4 & 1 & 1 & 1 \\ \hline 
	\end{tabular}}
	\label{tab1}
\end{table}

Many In-Memory Computing (IMC) operations are performed by specific arithmetic circuits such as adders and multipliers. Hence, optimized implementation of arithmetic circuits regarding speed, area, and energy consumption is crucial for efficient performance \cite{ref8, ref9, ref10, ref11, ref12, ref13, ref14, ref15}. Multiplication is a complex arithmetic operation requiring high circuit complexity to be appropriately implemented \cite{ref13}. There are many solutions to enhance the speed of the multiplication process in each stage of this operation, including Partial Product Generation (PPG), PPR, and Ripple Carry Adder (RCA) \cite{ref13}. Different sorts of multipliers, such as the Wallace tree \cite{ref13}, the Dadda tree \cite{ref14}, and the array multiplier \cite{ref15}, have been presented to enhance the performance of the multiplication.

Arithmetic circuits, such as full adder, counter, and compressor, are applied to implement the PPR tree in the multiplier \cite{ref15, ref16, ref17}. The compressor is also applied to implement multi-operand addition \cite{ref18, ref19}. It is essential to propose an efficient compressor design to improve the multiplier's speed, area, and energy consumption \cite{ref20}. Among the various types of compressors with different numbers of inputs and outputs, the 4:2 compressor has received much attention due to its regular structure and low complexity \cite{ref20, ref21, ref22}. Many researchers have presented different designs to reduce the complexity of this circuit and improve the delay, area, and energy consumption of the 4:2 compressor \cite{ref23, ref24, ref25}.

The implementation of the multiplier's PPR and RCA stages significantly affects its energy consumption and computational delay. Considering the importance of implementing multipliers for IMC in today's data-intensive applications, this article proposes a serial IMPLY-based 4:2 compressor cell, which is applied for the proposed multiplier's implementation to reduce the energy consumption and the number of computational steps of the multiplier. Furthermore, the combination of the PPR and RCA stages makes the maximum number of applied proposed 4:2 compressors and reduces the hardware complexity of the proposed IMPLY-based multiplier. The main contributions of this article are as follows:
\begin{itemize}
	\item Proposing a memristive IMPLY-based 4:2 compressor implementation algorithm; designing and implementing the proposed single-cell NAND-based 4:2 compressor in switch-free serial architecture;
	\item Outperforming IMPLY-based State-Of-the Art (SOA) in terms of the number of required memristors, computational steps, and energy consumption assessed by undergoing several detailed simulations;
	\item Presenting a new design for 4-bit and 8-bit multipliers based on the proposed 4:2 compressor's implementation algorithm;
	\item Reducing the multiplication tree and computing the RCA stage outputs by applying the proposed IMPLY-based serial half adder and single-cell NAND-based 4:2 compressor;
	\item Evaluating the proposed IMPLY-based crossbar array compatible multipliers regarding area and speed;
	\item Introducing equations to compute the proposed n-bit multiplier's number of memristors and computational steps.
	\end{itemize}

The remainder of this paper is arranged as follows: In Section \ref{sec2}, the literature of this research is reviewed. The design of the proposed 4:2 compressor, as well as the proposed 4×4 and 8×8 multipliers and their implementation are explicated in Section \ref{sec3}. Section \ref{sec4} presents the simulation results and evaluation of the proposed 4:2 compressor and multipliers regarding delay, area, and energy consumption. We compare the proposed designs with SOA in Section \ref{sec4}, concluding afterward in Section \ref{sec5}.

\section{Literature Review} \label{sec2}
\subsection{IMPLY-based Logic Circuits} \label{sec21}
FALSE and IMPLY logic functions form a complete computational logic structure. These functions can be applied to implement every single one of the basic logic gates and arithmetic circuits \cite{ref4, ref26}. Several basic Boolean logic gates implemented by IMPLY and FALSE logic operations are listed in Table \ref{tab2}. Memristive IMPLY-based arithmetic circuits are primarily implemented based on four design methods: serial \cite{ref9}, semi-serial \cite{ref28}, parallel \cite{ref27}, and semi-parallel \cite{ref29}. 

\begin{figure}[h]
	\centering
	\includegraphics[width=0.65\textwidth]{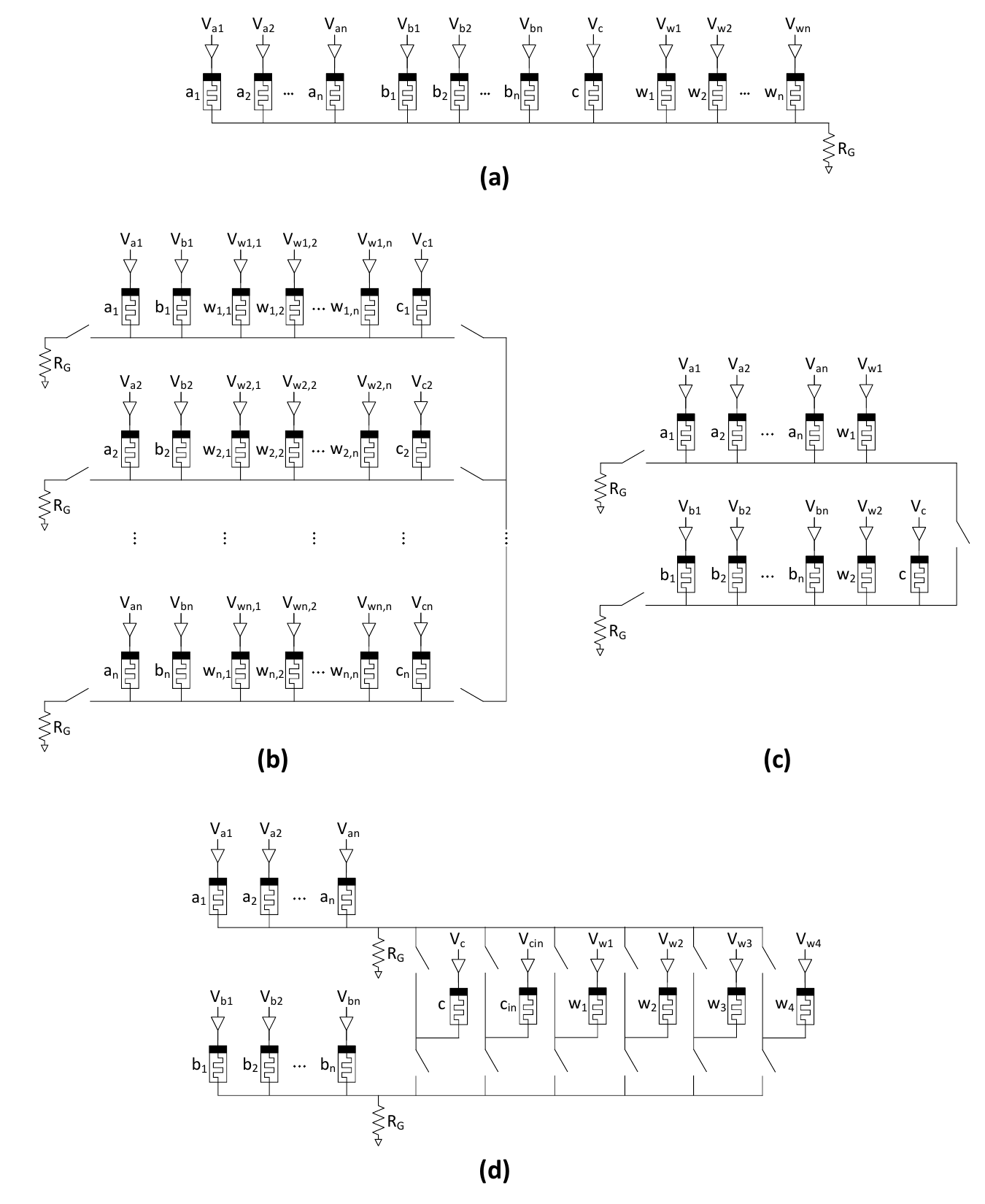}
	\caption{The IMPLY-based adder architectures: (a) serial, (b) parallel,(c) semi-parallel, and (d) semi-serial \cite{ref28}.}
	\label{fig2}
\end{figure}

The memristors in serial architecture are arranged serially in one row or column and connected to the ground through resistor $R_{G}$, depicted in Fig. \ref{fig2}(a) \cite{ref9}. Only one IMPLY or FALSE logic operation can be performed in each clock cycle, increasing the number of computational steps and execution time. Besides, this architecture has the least number of memristors and complexity among the existing design methods \cite{ref9}.

In the parallel architecture, as shown in Fig. \ref{fig2}(b), each section includes input memristors and work memristors of one bit of inputs, one $R_{G}$ resistor, and two CMOS switches to perform operations associated with the corresponding bit \cite{ref27}. The circuit elements of each section are connected in series, and the sections are connected in parallel. In this architecture, independent operations of bits can be performed simultaneously, decreasing step counts. However, in this architecture, the number of needed memristors for circuit implementation exceeds the serial one \cite{ref27}.

\begin{table}[h]
	\centering
	\caption{Implementation of basic Boolean logic gates using IMPLY and FALSE logic operations \cite{ref26}}
	\scalebox{1}{
		\begin{tabular}{|c|c|}
			\hline
			Logic gate & Equivalent IMPLY logic  \\ \hline
			NOT p & $p \to 0$ \\ \hline
			p OR q & $(p \to 0) \to q$ \\ \hline
			p NOR q & $((p \to 0) \to q) \to 0$ \\ \hline 
			p NAND q & $p \to (q \to 0)$ \\ \hline 
			p AND q & $(p \to (q \to 0)) \to 0$ \\ \hline 
			p XOR q & $(p \to q) \to ((q \to p) \to 0)$ \\ \hline 
	\end{tabular}}
	\label{tab2}
\end{table}

As illustrated in Fig. \ref{fig2}(c), in the semi-parallel architecture, there is a section for each input containing the input memristors, work memristors, and one resistor $R_{G}$ connected to the memristors via one CMOS switch \cite{ref29}. Also, the sections are connected via one CMOS switch to perform dependent operations between circuit inputs. This architecture has fewer memristors than the parallel design method and fewer computational steps than the serial architecture \cite{ref29}.

According to Fig. \ref{fig2}(d), in the semi-serial architecture, each section comprises the memristors of each input and one resistor $R_{G}$, and work memristors are shared between all sections applying CMOS switches \cite{ref28}. In this way, while the computational steps have been reduced by parallelizing independent operations of inputs, the number of memristors is somewhat decreased \cite{ref28}.

Among the mentioned architectures in \cite{ref26,ref28,ref29}, serial architecture is the only one compatible with the conventional crossbar array structure. Other architectures need CMOS switches to be added to the structure of the conventional crossbar array, called a modified crossbar array, so that the arithmetic and logic circuits designed in these architectures can be implemented and operated correctly \cite{ref28,ref29,ref43}.

\subsection{Full Adder}\label{sec22}
The full adder is an essential fundamental block of arithmetic circuits \cite{ref9, ref11, ref30}. This logic circuit has three single-bit and same valued inputs, $A$, $B$, and $C_{in}$, and produces two single-bit outputs, $Sum$ and $C_{out}$. The logic functions of the full adder’s outputs are expressed as (\ref{eq2}) and (\ref{eq3}) \cite{ref11}.
\begin{equation}
    Sum=A \oplus B \oplus C_{in}
    \label{eq2}
\end{equation}
\begin{equation}
    C_{out}=A \cdot B+C_{in}(A+B)
    \label{eq3}
\end{equation}
Several studies have investigated full adder circuits, and various memristive IMPLY-based full adders have been presented \cite{ref9, ref10, ref27, ref28, ref29}. In \cite{ref9}, a new algorithm for the computation of serial full adder is proposed, and an 8-bit adder based on the proposed full adder is presented. In the presented algorithm, by eliminating some repetitive operations, the computational steps of the full adder have been reduced to 22. Also, this full adder is implemented with five memristors by utilizing the input memristors to store the logical state of the outputs \cite{ref9}. Based on this algorithm, a serial n-bit adder is implemented in $22n$ computational steps by applying $2n+3$ memristors \cite{ref9}.

Currently, the best design of the n-bit RCA in the parallel architecture has been implemented with $4n+1$ memristors and executed in $5n+16$ computational steps \cite{ref27}. The authors improved the area occupation of this design by reusing input memristors as output memristors \cite{ref27}. Partitioning of the algorithm is another remarkable technique in which the memristor that stores the logical state of the carry bit is accessible by all parts \cite{ref27}.

Kaushik et al. proposed an n-bit Carry Select Adder (CSA) based on parallel topology by applying $\frac{19}{2}n+6$ memristors and executing in $3n+27$ computational steps \cite{rref44}. Moreover, the authors introduced and analyzed different-sized Conditional Carry Adders (CCAs) in parallel architecture \cite{rref44}.

In \cite{ref28}, the n-bit adder is implemented in semi-serial architecture, and the number of required memristors and computational steps are $2n+6$ and $10n+2$, respectively.

In \cite{ref29}, n-bit adder is presented in semi-parallel architecture. In the proposed algorithm, the independent operations of each input bit are executed in parallel. As a result, the computational steps have been reduced to $17n$. In addition, the number of memristors has been reduced to $2n+3$ by reusing input memristors as output ones \cite{ref29}.

\subsection{4:2 Compressor} \label{sec23}
\begin{figure}
	\centering
	\includegraphics[scale=0.25]{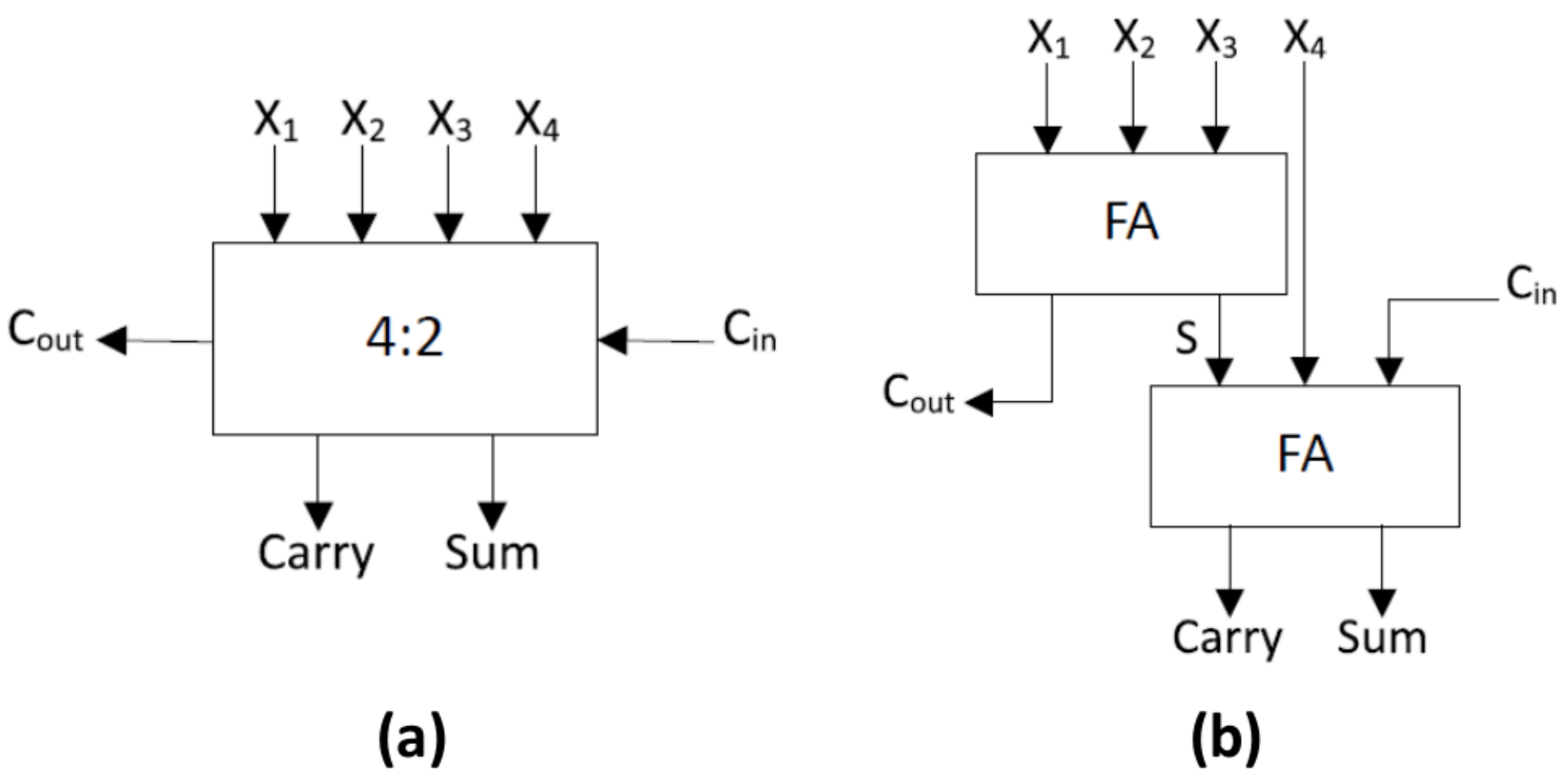}
	\caption{(a) 4:2 compressor module, and (b) conventional full adder based 4:2 compressor \cite{ref20}.}
	\label{fig3}
\end{figure}

\begin{figure}
	\centering
	\includegraphics[scale=0.27]{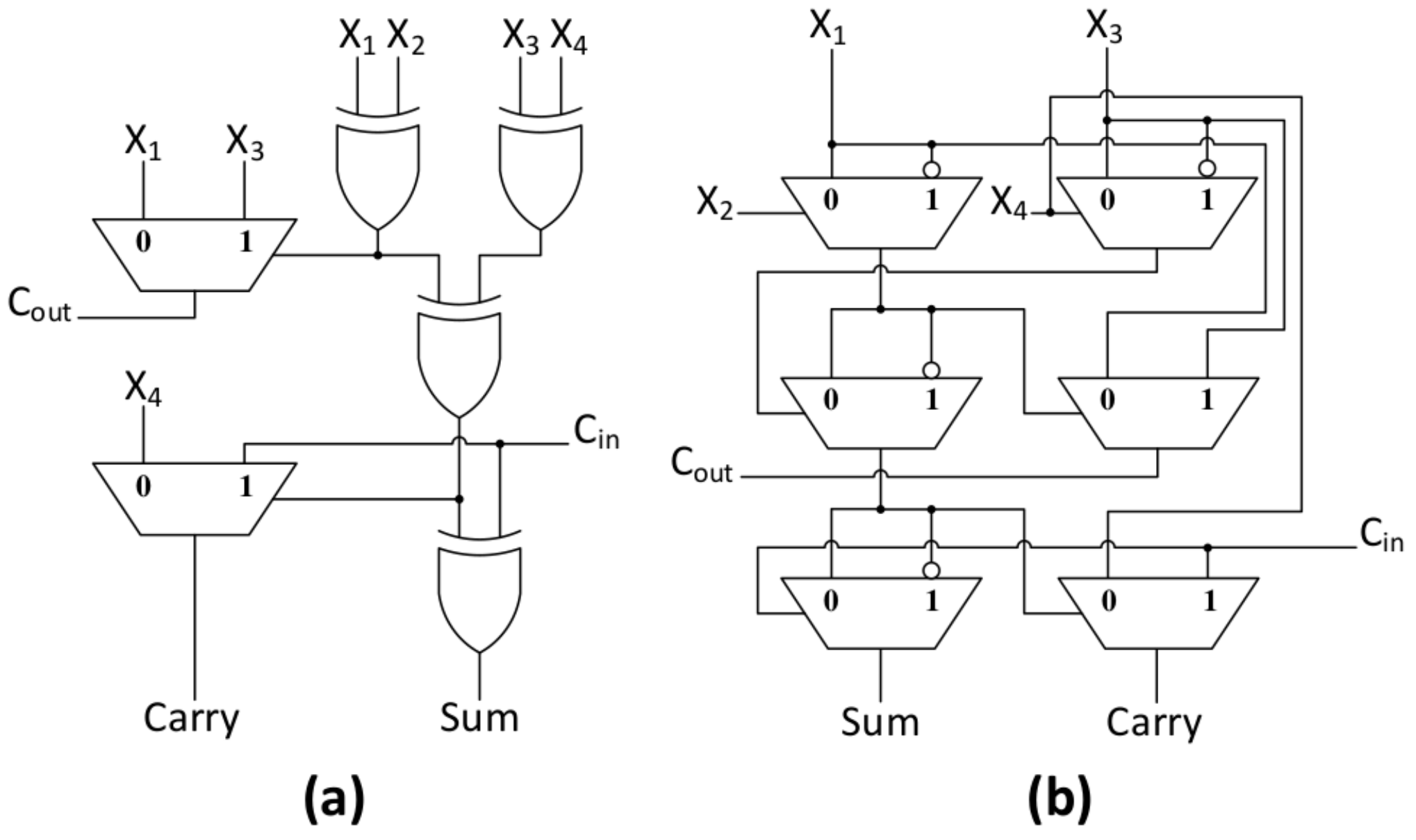}
	\caption{(a) XOR/MUX design of 4:2 compressor \cite{ref23}, and (b) 2:1 MUX based 4:2 compressor \cite{ref33}.}
	\label{fig4}
\end{figure}

Several compressors with different numbers of inputs and outputs have been presented, including 4:2, 5:2, 7:2, and 15:4 compressors \cite{ref19, ref21, ref22}. Weinberger introduced the 4:2 compressor with a regular structure and low complexity \cite{ref21}. It has received much attention for implementing large circuits such as multipliers, especially for the PPR tree \cite{ref31}. The 4:2 compressor has five single-bit inputs, $X_{1}$, $X_{2}$, $X_{3}$, $X_{4}$, and $C_{in}$, and produces three single-bit outputs, $C_{out}$, $Carry$, and $Sum$, depicted in Fig. \ref{fig3}(a). Most of the presented logic designs of the 4:2 compressor are based on (\ref{eq4}).
\begin{equation}
    X_{1}+X_{2}+X_{3}+X_{4}+C_{in}=Sum+2(Carry+C_{out})
    \label{eq4}
\end{equation}
One conventional design of the 4:2 compressor is based on two full adders connected serially. The schematic of this design is illustrated in Fig. \ref{fig3}(b). The Boolean functions of this design’s outputs, $C_{out}$, $Carry$, and $Sum$, are expressed as (\ref{eq5})–(\ref{eq7}), respectively \cite{ref20}.
\begin{equation}
    C_{out}=(X_{1} \oplus X_{2})\cdot X_{3}+\overline{(X_{1} \oplus X_{2})} \cdot X_{1}
    \label{eq5}
\end{equation}
\begin{equation}
   Carry=(X_{1} \oplus X_{2}\oplus X_{3}\oplus X_{4})\cdot C_{in}+ \overline{(X_{1} \oplus X_{2}\oplus X_{3}\oplus X_{4})}\cdot X_{4}
    \label{eq6}
\end{equation}
\begin{equation}
    Sum=X_{1} \oplus X_{2}\oplus X_{3}\oplus X_{4}\oplus C_{in}
    \label{eq7}
\end{equation}
One of the gate-level designs of the 4:2 compressor is based on four XOR gates and two 2:1 multiplexers, according to Fig. \ref{fig4}(a). Many previous works have investigated the XOR/MUX design of the 4:2 compressor due to its simple structure and low critical path delay \cite{ref20, ref23}. In \cite{ref32,ref33}, the 4:2 compressor is implemented based on six 2:1 multiplexers, depicted in Fig. \ref{fig4}(b). Another 4:2 compressor is designed based on the majority gate (MAJ), which is more complex than the previous three designs \cite{ref24}. 
At the circuit level, various implementations of 4:2 compressors have been presented based on existing and emerging technologies such as memristor \cite{ref23}, MOS Field Effect Transistor (MOSFET) \cite{ref34, ref35}, FinFET \cite{ref25}, Carbon Nanotube FET (CNFET) \cite{ref36}, and Quantum-dot Cellular Automata (QCA) \cite{ref37}.

In \cite{ref23}, a memristive 4:2 compressor is implemented by applying IMPLY logic. XOR/MUX design is used to implement the 4:2 compressor at the gate level (see Fig. \ref{fig4}). The applied XOR gate in this design is implemented based on a serial architecture in 9 computational steps and with 4 memristors \cite{ref23}. For the 2:1 multiplexer, \cite{ref23} presented a new algorithm that can be implemented in 7 computational steps by applying 4 memristors. Finally, the proposed algorithm of the 4:2 compressor has been implemented based on a serial architecture with 7 memristors, and the operations have been executed in 52 computational steps \cite{ref23}. The proposed serial implementation of the 4:2 compressor in \cite{ref23} has few memristors, but the computational steps are high. Implementing a 4:2 compressor based on parallel architecture has also been presented in \cite{ref23} to reduce the computational steps. 5 work memristors are used in this design, and FALSE operations can be executed simultaneously in one step. According to the presented algorithm in \cite{ref23}, two XOR gates are first parallelized, so the computational steps of these two gates are reduced from 18 to 9. Then, the operations of one XOR gate and 2:1 multiplexer are executed simultaneously in 9 computational steps, and the $C_{out}$ output of the 4:2 compressor is computed. Finally, at the third stage of the operation, one XOR gate and one 2:1 multiplexer are parallelized, and the other outputs of the 4:2 compressor ($Sum$ and $Carry$) are calculated in step 26. Consequently, the XOR/MUX design of the 4:2 compressor based on parallel architecture is implemented with 11 memristors and in 26 computational steps \cite{ref23}.

\subsection{Multiplier}\label{sec24}
Several studies have been conducted to accelerate multiplication, mainly focusing on PPR and speeding up addition \cite{ref13, ref14}. The multiplication process generally consists of three phases: 1) PPG, 2) PPR, and 3) RCA. In the first phase, partial products are generated by applying AND gates. The second phase involves using adders and compressors to reduce partial products. The two final $Sum$ and $Carry$ rows obtained from the second phase are added in the third phase, and the final product is computed \cite{ref12}.

Wallace and Dadda multipliers are some of the multipliers based on conventional multiplication architecture. Half adder, full adder, and 4:2 compressor are common arithmetic circuits applied to implement Wallace and Dadda multipliers \cite{ref13, ref14}.

Another well-known multiplier is the array multiplier, which consists of several processing units in the second phase \cite{ref15, ref38}. Each of these processing units comprises AND (or NAND) gates and a 2:2 counter (half adder) or 3:2 counter (full adder), which is used in signed and unsigned array multipliers. The $n \times n$ array multiplier is implemented with $n^2-n$ number of processing units \cite{ref15, ref38}.

The literature presents various memristive IMPLY-based nonserial multipliers. In \cite{ref28}, the n-bit multiplier is implemented using $\lceil \frac{n}{2} \rceil$ numbers of $2n-1$ bit adders based on an IMPLY logic design with semi-serial architecture. 

In \cite{ref39}, a parallel 8-bit Dadda multiplier is implemented, applying full adders and half adders. Guckert et al. presented a parallel array multiplier in \cite{ref15}, in which the memristors of AND gates in the first phase are reused for operations of the second phase. However, the presented parallel multipliers in \cite{ref39, ref15} have high complexity and overhead due to the CMOS switches and $R_{G}$ resistors applied to each section of the circuits (see Fig. \ref{fig2}). 

On the other hand, due to the placement of CMOS switches, a modified crossbar array is needed for these semi-serial and parallel circuits to be applied  \cite{ref28, ref39, ref15, ref43}. Our preliminary purpose is to ensure the circuits' compatibility with the conventional crossbar array structure; hence, the proposed arithmetic circuits are implemented in serial architecture. Therefore, the multipliers in \cite{ref28, ref39, ref15} are not examined and evaluated in the following sections.

\section{Proposed IMPLY-based Circuits}\label{sec3}
\subsection{Proposed 4:2 Compressor}\label{sec31}
\begin{figure}[h]
	\centering
	\includegraphics[scale=0.3]{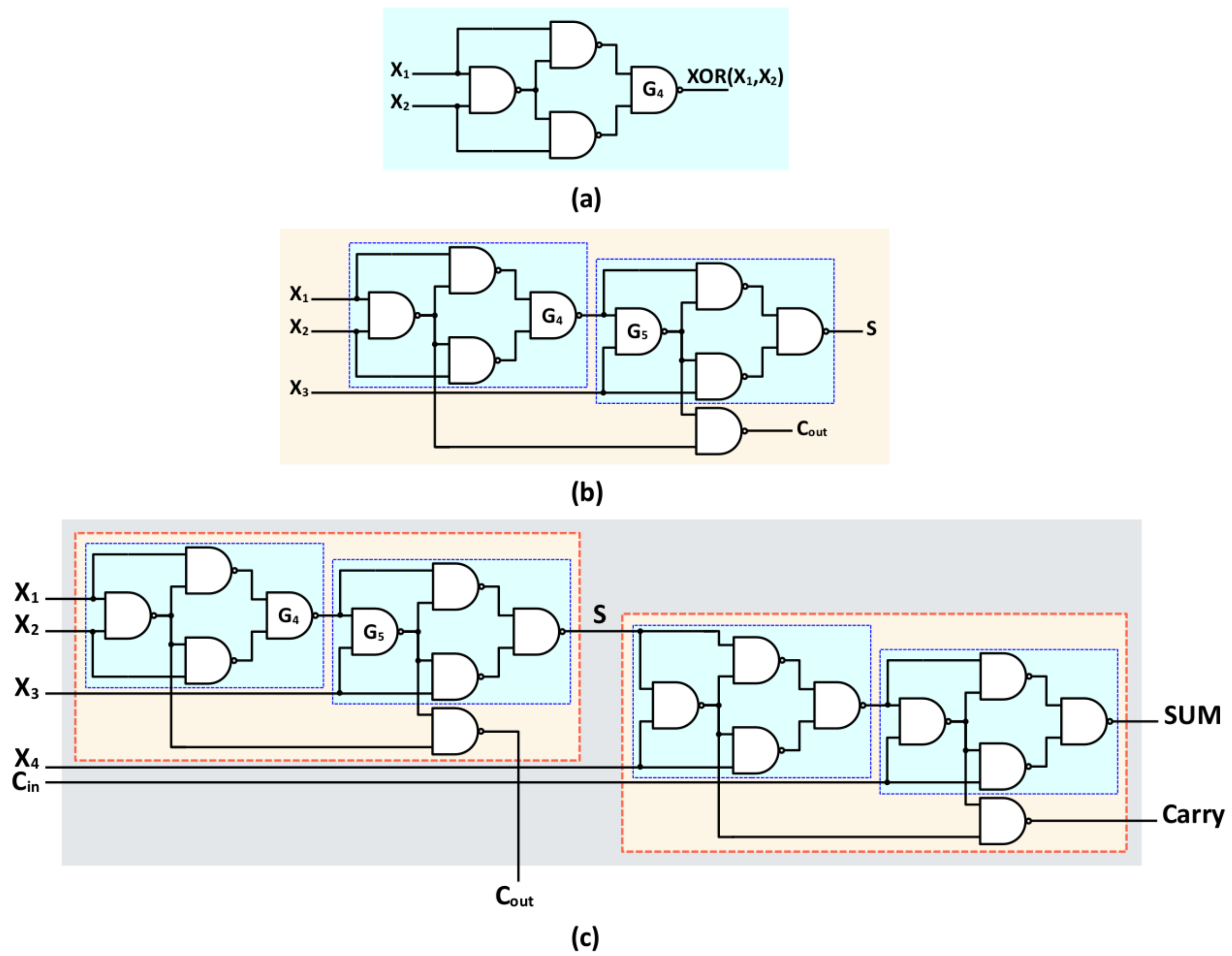}
	\caption{(a) Design of a NAND-based XOR gate, (b) design of an XOR-based full adder based on NAND gates, and (c) NAND-based 4:2 compressor}
	\label{fig5}
\end{figure}

As explained in Section \ref{sec2}, applying the XOR gate, multiplexer, and MAJ function units are among the common designs for implementing the 4:2 compressor. Every logic function can be designed and implemented by applying NAND or NOR, universal logic gates. Three computational steps are needed to implement the IMPLY-based NAND gate while implementing the IMPLY-based NOR gate requires five computational steps with the same number of memristors. Due to this advantage, the 4:2 compressor is designed entirely with a two-input NAND gate, depicted in Fig. \ref{fig5}(c). The XOR gate is implemented with four NAND gates, in the form of (\ref{eq8}), in the first step of designing the 4:2 compressor.
\begin{equation}
    G_{4}=X_{1} \oplus X_{2}=X_{1}\overline{X_{2}}+\overline{X_{1}}X_{2}=(X_{1}+X_{2})\cdot (\overline{X_{1}}+\overline{X_{2}})=(X_{1}+X_{2})\cdot \overline{X_{1}X_{2}}=\overline{\overline{(X_{1}\cdot \overline{X_{1}X_{2}})}\cdot \overline{(X_{2}\cdot \overline{X_{1}X_{2}}})}
    \label{eq8}
  \end{equation}
 The NAND-based XOR gate, with its inputs ($X_{1}$ and $X_{2}$) and output ($G_{4}$), is shown in Fig. \ref{fig5}(a). In the next step, the NAND-based full adder cell is implemented. The logic functions of the full adder cell were examined in Section \ref{sec2}. According to Fig. \ref{fig5}(b), two NAND-based XOR gates are connected in series to implement the full adder’s $Sum$ output with its logic function in (\ref{eq9}).
 \begin{equation}
 S=\overline{(\overline{(\overline{(G_{4}\cdot X_{3})}\cdot G_{4})}\cdot \overline{(\overline{(G_{4}\cdot X_{3})}\cdot X_{3})})}
 \label{eq9}
 \end{equation}
 The $C_{out}$ output of the NAND-based full adder is expressed as (\ref{eq10}), applying the Boolean algebra and De Morgan’s law. A part of the $C_{out}$ output function, $G_{4}=(X_{1} \oplus X_{2})$, is already implemented. Consequently, the full adder cell is designed by applying 9 NAND gates, depicted in Fig. \ref{fig5}(b). In the last step, two NAND-based full adders are connected in series to implement the 4:2 compressor. The IMPLY-based 4:2 compressor is implemented with 18 NAND gates, according to Fig. \ref{fig5}(c). The first full adder’s $C_{out}$ output is the $C_{out}$ output of the 4:2 compressor, and the outputs of the second full adder are the $Carry$ and $SUM$ outputs of the 4:2 compressor. The logic functions of the $C_{out}$, $Carry$, and $SUM$ outputs of the NAND-based 4:2 compressor are presented as (\ref{eq10})–(\ref{eq12}), respectively.
 \begin{equation}
     C_{out}=\overline{(\overline{X_{1}\cdot X_{2})}\cdot \overline{(G_{4}\cdot X_{3}})}
     \label{eq10}
 \end{equation}
 \begin{equation}
     Carry= \\
     \overline{(\overline{(\overline{(\overline{(\overline{(S\cdot X_{4})} \cdot S)}\cdot \overline{(\overline{(S \cdot X_{4})}\cdot X_{4})})}\cdot C_{in})}\cdot \overline{(S\cdot X_{4})})}= \overline{(\overline{((S\oplus X_{4})\cdot C_{in})}\cdot \overline{(S\cdot X_{4})})}
     \label{eq11}
 \end{equation}
 \begin{equation}
     SUM=\overline{(\overline{(\overline{((S\oplus X_{4})\cdot C_{in})}\cdot (S\oplus X_{4}))}\cdot \overline{(\overline{((S\oplus X_{4})\cdot C_{in})}\cdot C_{in}}))}
     \label{eq12}
 \end{equation}

\begin{figure}[t]
	\centering
	\includegraphics[scale=0.5]{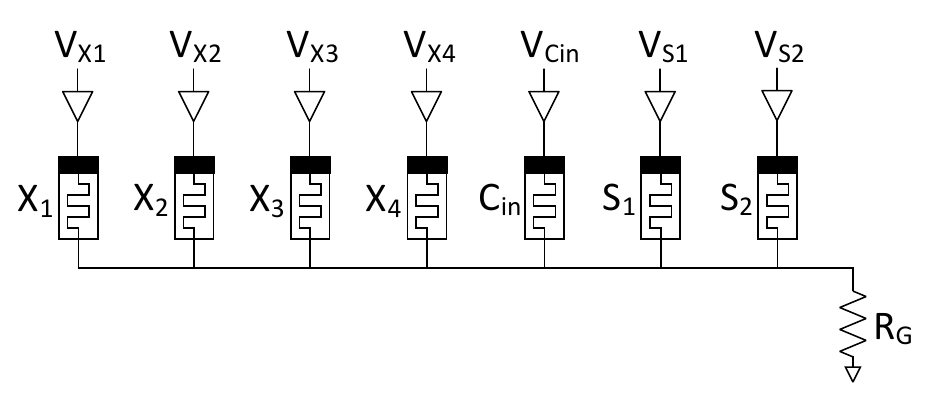}
	\caption{Proposed memristive 4:2 compressor's implementation in serial architecture.}
	\label{fig6}
\end{figure}

 The proposed memristive 4:2 compressor is implemented based on the IMPLY logic with serial architecture at the circuit level, as depicted in Fig. \ref{fig6}. Five memristors are applied to store the inputs’ logic states of the proposed 4:2 compressor, i.e., $X_{1}$, $X_{2}$, $X_{3}$, $X_{4}$, and $C_{in}$. Also, two work memristors, $S_{1}$ and $S_{2}$, are applied to perform IMPLY operations and maintain the results of operations. The proposed circuit’s implementation algorithm is presented in Table \ref{tab3}. The proposed algorithm is the 4:2 compressor’s implementation algorithm as a cell design, and each full adder is not considered an individual component in this algorithm.
 
 In the FALSE operations of Table \ref{tab3}, $V_{RESET}$ is applied across the intended memristor. For instance, in the second step of the proposed 4:2 compressor’s algorithm, $V_{RESET}$ is applied across the memristor $S_{2}$. In other steps of the proposed 4:2 compressor’s algorithm, $V_{COND}$ and $V_{SET}$ voltages are applied across the involved memristors of IMPLY operation. In the third step of the proposed 4:2 compressor’s algorithm, as an example, $V_{COND}$ and $V_{SET}$ are applied to memristors $X_{2}$ and $S_{1}$, respectively. 

\begin{table}
	\centering
	\caption{Algorithm of NAND-based 4:2 compressor with serial implementation}
	\scalebox{0.9}{
		\begin{tabular}{|c|c|c|}
			\hline
			\textbf{Step} & \textbf{Operation} & \textbf{Equivalent Logic} \\ \hline
			\textbf{1} & $S_{1}=0$ & FALSE($S_{1}$) \\ \hline
			\textbf{2} & $S_{2}=0$ & FALSE($S_{2}$) \\ \hline
			\textbf{3} & $(X_{2} \to 0) \equiv (X_{2} \to S_{1})=S_{1}^{'}$ & $S_{1}^{'} = \overline{X_{2}}$ \\ \hline
			\textbf{4} & $(X_{1} \to S_{1}^{'}) = S_{1}^{''}$ & $S_{1}^{''} = \overline{X_{1}} + \overline{X_{2}}$ \\ \hline
			\textbf{5} & $(X_{1} \to 0) \equiv (X_{1} \to S_{2})=S_{2}^{'}$ & $S_{2}^{'} = \overline{X_{1}}$ \\ \hline
			\textbf{6} & $(S_{2}^{'} \to X_{2}) = X_{2}^{'}$ & $X_{2}^{'} = X_{1} + X_{2}$ \\ \hline
			\textbf{7} & $S_{2}=0$ & FALSE($S_{2}$)  \\ \hline
			\textbf{8} & $(S_{1}^{''} \to 0) \equiv (S_{1}^{''} \to S_{2})=S_{2}^{'}$ & $S_{2}^{'} = X_{1} \cdot X_{2}$ \\ \hline
			\textbf{9} & $(X_{2}^{'} \to S_{2}^{'}) = S_{2}^{''}$ & $S_{2}^{''} = \overline{X_{1}} \cdot \overline{X_{2}} + X_{1} \cdot X_{2}$ \\ \hline
			\textbf{10} & $X_{2}=0$ & FALSE($X_{2}$)  \\ \hline
			\textbf{11} & $(S_{2}^{''} \to 0) \equiv (S_{2}^{''} \to X_{2}) = X_{2}^{'}$ & $X_{2}^{'} = (X_{1} + X_{2}) \cdot (\overline{X_{1}} + \overline{X_{2}}) = (X_{1} \oplus X_{2}) = \bm{G_{4}}$ \\ \hline
			\textbf{12} & $(X_{3} \to S_{2}^{''}) = S_{2}^{'''}$ & $S_{2}^{'''} = \overline{X_{3}} + \overline{X_{1}} \cdot \overline{X_{2}} + X_{1} \cdot X_{2} = \bm{G_{5}}$ \\ \hline
			\textbf{13} & $X_{1}=0$ & FALSE($X_{1}$) \\ \hline
			\textbf{14} & $(S_{2}^{'''} \to 0) \equiv (S_{2}^{'''} \to X_{1}) = X_{1}^{'}$ & $X_{1}^{'} = \overline{G_{5}}$ \\ \hline
			\textbf{15} & $(S_{1}^{''} \to X_{1}^{'}) = X_{1}^{''}$ & \cellcolor{tablecolor}$X_{1}^{''} = X_{1} \cdot X_{2} + \overline{G_{5}} = (X_{1} \cdot X_{2}) + X_{3} \cdot (X_{1} \oplus X_{2}) = \bm{\textcolor{red}{C_{out}}}$ \\ \hline
			
			\textbf{16} & $S_{1}=0$ & FALSE($S_{1}$) \\ \hline
			\textbf{17} & $(X_{2}^{'} \to 0) \equiv (X_{2}^{'} \to S_{1}) = S_{1}^{'}$ & $S_{1}^{'} = \overline{G_{4}}$ \\ \hline
			\textbf{18} & $(S_{1}^{'} \to X_{3}) = X_{3}^{'}$ & $X_{3}^{'} = X_{3} + G_{4}$ \\ \hline
			\textbf{19} & $S_{1}=0$ & FALSE($S_{1}$)  \\ \hline
			\textbf{20} & $(S_{2}^{'''} \to 0) \equiv (S_{2}^{'''} \to S_{1}) = S_{1}^{'}$ & $S_{1}^{'} = \overline{G_{5}}$ \\ \hline
			\textbf{21} & $(X_{3}^{'} \to S_{1}^{'}) = S_{1}^{''}$ & $S_{1}^{''} = \overline{G_{4}} \cdot \overline{X_{3}} + \overline{G_{5}} $ \\ \hline
			\textbf{22} & $X_{3}=0$ & FALSE($X_{3}$)  \\ \hline
			\textbf{23} & $(S_{1}^{''} \to 0) \equiv (S_{1}^{''} \to X_{3}) = X_{3}^{'}$ & $X_{3}^{'} = (X_{3} + G_{4}) \cdot G_{5} = (X_{1} \oplus X_{2} \oplus X_{3}) = \bm{S}$ \\ \hline
			\textbf{24} & $(X_{4} \to S_{1}^{''}) = S_{1}^{'''}$ & $S_{1}^{'''} = \overline{X_{4}} + \overline{S}$ \\ \hline
			\textbf{25} & $X_{2}=0$ & FALSE($X_{2}$)  \\ \hline
			\textbf{26} & $(X_{3}^{'} \to 0) \equiv (X_{3}^{'} \to X_{2}) = X_{2}^{'}$ & $X_{2}^{'} = \overline{S}$ \\ \hline
			\textbf{27} & $(X_{2}^{'} \to X_{4}) = X_{4}^{'}$ & $X_{4}^{'} = S + X_{4}$ \\ \hline
			\textbf{28} & $X_{2}=0$ & FALSE($X_{2}$)  \\ \hline
			\textbf{29} & $(S_{1}^{'''} \to 0) \equiv (S_{1}^{'''} \to X_{2}) = X_{2}^{'}$ & $X_{2}^{'} = X_{4} \cdot S $ \\ \hline
			\textbf{30} & $(X_{4}^{'} \to X_{2}^{'}) = X_{2}^{''}$ & $X_{2}^{''} = \overline{X_{4}} \cdot \overline{S} + X_{4} \cdot S = \overline{X_{1} \oplus X_{2} \oplus X_{3} \oplus X_{4}}$ \\ \hline
			\textbf{31} & $X_{4}=0$ & FALSE($X_{4}$)  \\ \hline
			\textbf{32} & $(X_{2}^{''} \to 0) \equiv (X_{2}^{''} \to X_{4}) = X_{4}^{'}$ & $X_{4}^{'} = (S + X_{4}) \cdot (\overline{X_{4}} + \overline{S}) = X_{1} \oplus X_{2} \oplus X_{3} \oplus X_{4}$ \\ \hline
			\textbf{33} & $(C_{in} \to X_{2}^{''}) = X_{2}^{'''}$ & $X_{2}^{'''} = \overline{C_{in}} + \overline{X_{1} \oplus X_{2} \oplus X_{3} \oplus X_{4}}$ \\ \hline
			\textbf{34} & $X_{3}=0$ & FALSE($X_{3}$)  \\ \hline
			\textbf{35} & $(S_{1}^{'''} \to 0) \equiv (S_{1}^{'''} \to X_{3}) = X_{3}^{'}$ & $X_{3}^{'} = X_{4} \cdot S$ \\ \hline
			\textbf{36} & $(X_{2}^{'''} \to X_{3}^{'}) = X_{3}^{''}$ & \cellcolor{tablecolor}$X_{3}^{''} = X_{4} \cdot S + C_{in} \cdot X_{1} \oplus X_{2} \oplus X_{3} \oplus X_{4} = \bm{\textcolor{red}{Carry}}$ \\ \hline
			\textbf{37} & $S_{2}=0$ & FALSE($S_{2}$)  \\ \hline
			\textbf{38} & $(X_{4}^{'} \to 0) \equiv (X_{4}^{'} \to S_{2}) = S_{2}^{'}$ & $S_{2}^{'} = \overline{X_{1} \oplus X_{2} \oplus X_{3} \oplus X_{4}}$ \\ \hline
			\textbf{39} & $(S_{2}^{'} \to C_{in}) = C_{in}^{'}$ & $C_{in}^{'} = C_{in} + X_{1} \oplus X_{2} \oplus X_{3} \oplus X_{4}$ \\ \hline
			\textbf{40} & $S_{1}=0$ & FALSE($S_{1}$)  \\ \hline
			\textbf{41} & $(X_{2}^{'''} \to 0) \equiv (X_{2}^{'''} \to S_{1}) = S_{1}^{'}$ & $S_{1}^{'} = C_{in} \cdot X_{1} \oplus X_{2} \oplus X_{3} \oplus X_{4}$ \\ \hline
			\textbf{42} & $(C_{in}^{'} \to S_{1}^{'}) = S_{1}^{''}$ & $S_{1}^{''} = \overline{C_{in}} \cdot \overline{X_{1} \oplus X_{2} \oplus X_{3} \oplus X_{4}} + C_{in} \cdot X_{1} \oplus X_{2} \oplus X_{3} \oplus X_{4}$ \\ \hline
			\textbf{43} & $C_{in}=0$ & FALSE($C_{in}$)  \\ \hline
			\textbf{44} & $(S_{1}^{''} \to 0) \equiv (S_{1}^{''} \to C_{in}) = C_{in}^{'}$ & \cellcolor{tablecolor}$C_{in}^{'} = (C_{in} + X_{1} \oplus X_{2} \oplus X_{3} \oplus X_{4}) \cdot (\overline{C_{in}} + \overline{X_{1} \oplus X_{2} \oplus X_{3} \oplus X_{4}}) $ \\
			& & \cellcolor{tablecolor}$= X_{1} \oplus X_{2} \oplus X_{3} \oplus X_{4} \oplus C_{in} = \bm{\textcolor{red}{Sum}}$ \\ \hline
	\end{tabular}}
	\label{tab3}
\end{table}

 As mentioned earlier, implementing an IMPLY-based NAND gate in the serial architecture requires 3 computational steps. The proposed 4:2 compressor is implemented with 18 NAND gates serially. Therefore, 54 computational steps are needed to calculate the proposed IMPLY-based circuit with serial architecture. As explained before, the result of the IMPLY function, $p \to q \equiv \overline{p}+q$, is stored in the memristor $q$, and its input logic state is cleared. The IMPLY-based function of $G_{4}$ is presented in (\ref{eq13}), which is converted to (\ref{eq14}) by simplifying the operation.
 \begin{equation}
     G_{4}=[(X_{1}\to (X_{2} \to 0)) \to (X_{1} \to 0)] \to [((X_{1}\to (X_{2} \to 0))\to (X_{2} \to 0))\to 0]
     \label{eq13}     
 \end{equation}
 \begin{equation}
     G_{4}=\\
     \bm{\textcolor{red}{[}}((X_{1} \to 0)\to X_{2})\to ((X_{1}\to (X_{2}\to 0))\to 0) \bm{\textcolor{red}{]}} \to 0 \equiv \bm{\textcolor{red}{\alpha}} \to 0
     \label{eq14} 
 \end{equation}
 The IMPLY-based logic expression $\alpha$ equals the left statement of the last IMPLY operation of (\ref{eq14}), i.e., $\alpha \to 0$. Therefore, its corresponding memristor is only applied as an input, and its logic state can be saved for subsequent computations. The computation of IMPLY-based logic expression $\alpha$ is repeated in (\ref{eq15}) for the $G_{5}$ gate.
 \begin{equation}
     G_{5}= \\
     X_{3} \to \bm{\textcolor{red}{[}}((X_{1} \to 0)\to X_{2})\to ((X_{1}\to (X_{2}\to 0))\to 0)\bm{\textcolor{red}{]}} \equiv X_{3} \to \bm{\textcolor{red}{\alpha}}
     \label{eq15}
 \end{equation}
 $\alpha$ equals the right statement of the last IMPLY operation of (\ref{eq15}). Hence, at the beginning of the computation of this IMPLY operation, the logic state of the $\alpha$’s corresponding memristor is applied as an input, and at the end of the operation, the result of the computation is saved in it as an output. The serial implementation algorithm of the proposed 4:2 compressor is given in Table \ref{tab3}. In the proposed algorithm, the execution of repetitive operations is avoided as much as possible to decrease the computational steps. According to Table \ref{tab3}, the computation of $\alpha$ is executed up to step 9, and the result is saved in the work memristor, $S_{2}$. Then, the operation $\alpha \to 0$ of $G_{4}$ is executed in steps 10 and 11, and its result is saved in memristor $X_{2}$ at the $11^{th}$ computational step. The stored value of $\alpha$ in memristor $S_{2}$ is maintained till step 11, so it can be applied to compute the last IMPLY operation of $G_{5}$ in step 12. The output logic state of $G_{5}$ can be computed only in one step, the $12^{th}$ computational step, by applying $S_{2}$ in the IMPLY operation of this step. The $G_{5}$ result is saved in memristor $S_{2}$. Thus, $G_{5}$’s output, implemented based on five serial NAND gates, is obtained in 12 computational steps instead of 15. This method of overlapping and preventing the execution of repetitive operations is applied in the entire implementation algorithm of the proposed 4:2 compressor, and the number of computational steps is decreased from 54 to 44. According to the algorithm of Table \ref{tab3}, $C_{out}$, $Carry$, and $SUM$ outputs are computed in steps 15, 36, and 44, respectively. The IMPLY-based NAND gate’s output is stored in the work memristor. Hence, in implementing the NAND-based circuit, one work memristor is applied for each NAND gate to store the result of the gate operation, which increases the number of required memristors. The outputs of the proposed 4:2 compressor, each equivalent to the output of a NAND gate (see Fig. \ref{fig5}), are stored in work memristors. Therefore, in the case of implementing a larger circuit such as a multiplier, based on the proposed 4:2 compressor, the number of work memristors and, as a result, the total number of memristors of the circuit increases. The implementation algorithm of the proposed 4:2 compressor is written in such a way that the input memristors that remain unused after the completion of the first full adder’s operation, especially $X_{1}$, $X_{2}$, and $X_{3}$, are reused as work memristors to solve this issue. According to the algorithm in Table \ref{tab3}, the proposed 4:2 compressor’s $C_{out}$, $Carry$, and $SUM$ are stored in the input memristors $X_{1}$, $X_{3}$, and $C_{in}$, respectively. Thus, the proposed memristive NAND-based 4:2 compressor is implemented in serial architecture with 7 memristors and in 44 computational steps. 

 \subsection{Proposed Multipliers}\label{sec32}
 The conventional designs for the PPR tree of multiplier were studied in Section \ref{sec2}, including the Wallace tree, the Dadda tree, and the array multiplier. Conventionally, the multiplication process comprises three phases of PPG, PPR, and RCA to obtain the multiplication result \cite{ref40}. The proposed multiplier takes a distinct approach to PPR, as the last two phases are merged until the final product is attained. The reason is that the implementation and the execution of operations are done serially in the proposed design. Therefore, gate-level parallelization is pointless. The proposed multiplier is implemented with the maximum number of proposed 4:2 compressors by merging these two phases. In the following subsections, the proposed 4-bit and 8-bit multipliers with their memristive IMPLY-based serial circuit implementation are investigated in each phase.

 \subsubsection{4$\times$4 Multiplier} \label{sec321} 
 At Phase \Romannum{1} of the proposed multiplier, partial products are generated by multiplying the inputs bit by bit using 16 AND gates. Each input bit requires one memristor to store its logic state. So, 8 input memristors are applied for the 4-bit multiplier. The IMPLY-based serial AND gate is implemented with 4 memristors (2 input memristors+2 work memristors). The execution of the AND gate takes 5 computational steps, according to the Boolean function of basic gates based on IMPLY and FALSE operations presented in Table \ref{tab2}. Based on the algorithm of the AND gate, the logic state of its output is stored in the work memristor to keep the logic state of the inputs. Hence, one of the two work memristors stores the generated partial product at the end of the AND gate operation. The other work memristor remains unused and can be shared between all AND gates. Table \ref{tab4} presents the execution order of AND gates with each gate's involved input, work, and output memristors for Phase \Romannum{1} of the 4-bit multiplier. According to this table, the output of the $4^{th}$ AND gate is partial product $a_{0}b_{0}$, equal to the LSB bit of the final product, i.e., $y_{0}$. After generating the last partial product of $a_{j}b_{0} (j=0, 1, 2, 3)$, the memristor $b_{0}$ remains unused. $a_{0}b_{0}$ is generated as the last partial product of row $a_{j}b_{0} (j=0, 1, 2, 3)$, so it can be stored in the memristor $b_{0}$. Similarly, the last partial product generated in rows $a_{j}b_{1}$ and $a_{j}b_{2} (j=0, 1, 2, 3)$ are stored in memristors $b_{1}$ and $b_{2}$, respectively, to reduce the number of memristors. Also, the last row’s partial products, $a_{0}b_{3}$, $a_{1}b_{3}$, $a_{2}b_{3}$, and $a_{3}b_{3}$, are stored in memristors $a_{0}$, $a_{1}$, $a_{2}$, and $a_{3}$, respectively. According to Table \ref{tab4}, the outputs of the $8^{th}$ and $12^{th}$-$16^{th}$ AND gates are stored in the input memristors $b_{1}$, $b_{2}$, $a_{0}$, $a_{1}$, $a_{2}$, and $a_{3}$, respectively. Finally, 18 memristors (8 input and 10 work memristors) and 80 computational steps are required to implement Phase \Romannum{1} of the proposed 4-bit multiplier.
 
 \begin{table}
	\centering
	\caption{The execution order of AND gates with involved memristors (Phase \Romannum{1} of 4$\times$4 multiplier)}
	\scalebox{0.9}{
		\begin{tabular}{|c|c|c|c|c|}
			\hline
			 & \textbf{AND} & \textbf{Input} & \textbf{Work} & \textbf{Output}  \\
              & \textbf{Operation} & \textbf{Memristors} & \textbf{Memristors} & \textbf{Memristor} \\ \hline 
            \textbf{1} & $a_{1} \cdot b_{0}$ & $a_{1}$, $ b_{0}$ & $S_{1}$, $S_{2}$ & $S_{2}$  \\ \hline
            \textbf{2} & $a_{2} \cdot b_{0}$ & $a_{2}$, $ b_{0}$ & $S_{1}$, $S_{3}$ & $S_{3}$  \\ \hline
            \textbf{3} & $a_{3} \cdot b_{0}$ & $a_{3}$, $ b_{0}$ & $S_{1}$, $S_{4}$ & $S_{4}$  \\ \hline
            \textbf{4} & $a_{0} \cdot b_{0}$ & $a_{0}$, $ b_{0}$ & $S_{1}$ & $b_{0}$$\bm{=y_{0}}$  \\ \hline
            \textbf{5} & $a_{0} \cdot b_{1}$ & $a_{0}$, $ b_{1}$ & $S_{1}$, $S_{5}$ & $S_{5}$  \\ \hline
            \textbf{6} & $a_{1} \cdot b_{1}$ & $a_{1}$, $ b_{1}$ & $S_{1}$, $S_{6}$ & $S_{6}$  \\ \hline
            \textbf{7} & $a_{2} \cdot b_{1}$ & $a_{2}$, $ b_{1}$ & $S_{1}$, $S_{7}$ & $S_{7}$  \\ \hline
            \textbf{8} & $a_{3} \cdot b_{1}$ & $a_{3}$, $ b_{1}$ & $S_{1}$ & $b_{1}$  \\ \hline
            \textbf{9} & $a_{1} \cdot b_{2}$ & $a_{1}$, $ b_{2}$ & $S_{1}$, $S_{8}$ & $S_{8}$  \\ \hline
            \textbf{10} & $a_{2} \cdot b_{2}$ & $a_{2}$, $ b_{2}$ & $S_{1}$, $S_{9}$ & $S_{9}$  \\ \hline
            \textbf{11} & $a_{3} \cdot b_{2}$ & $a_{3}$, $ b_{2}$ & $S_{1}$, $S_{10}$ & $S_{10}$  \\ \hline
            \textbf{12} & $a_{0} \cdot b_{2}$ & $a_{0}$, $ b_{2}$ & $S_{1}$ & $b_{2}$  \\ \hline
            \textbf{13} & $a_{0} \cdot b_{3}$ & $a_{0}$, $ b_{3}$ & $S_{1}$ & $a_{0}$  \\ \hline
            \textbf{14} & $a_{1} \cdot b_{3}$ & $a_{1}$, $ b_{3}$ & $S_{1}$ & $a_{1}$  \\ \hline
            \textbf{15} & $a_{2} \cdot b_{3}$ & $a_{2}$, $ b_{3}$ & $S_{1}$ & $a_{2}$  \\ \hline
            \textbf{16} & $a_{3} \cdot b_{3}$ & $a_{3}$, $ b_{3}$ & $S_{1}$ & $a_{3}$  \\ \hline
	\end{tabular}}
	\label{tab4}
\end{table}

 \begin{figure}[!h]
	\centering
	\includegraphics[scale=0.3]{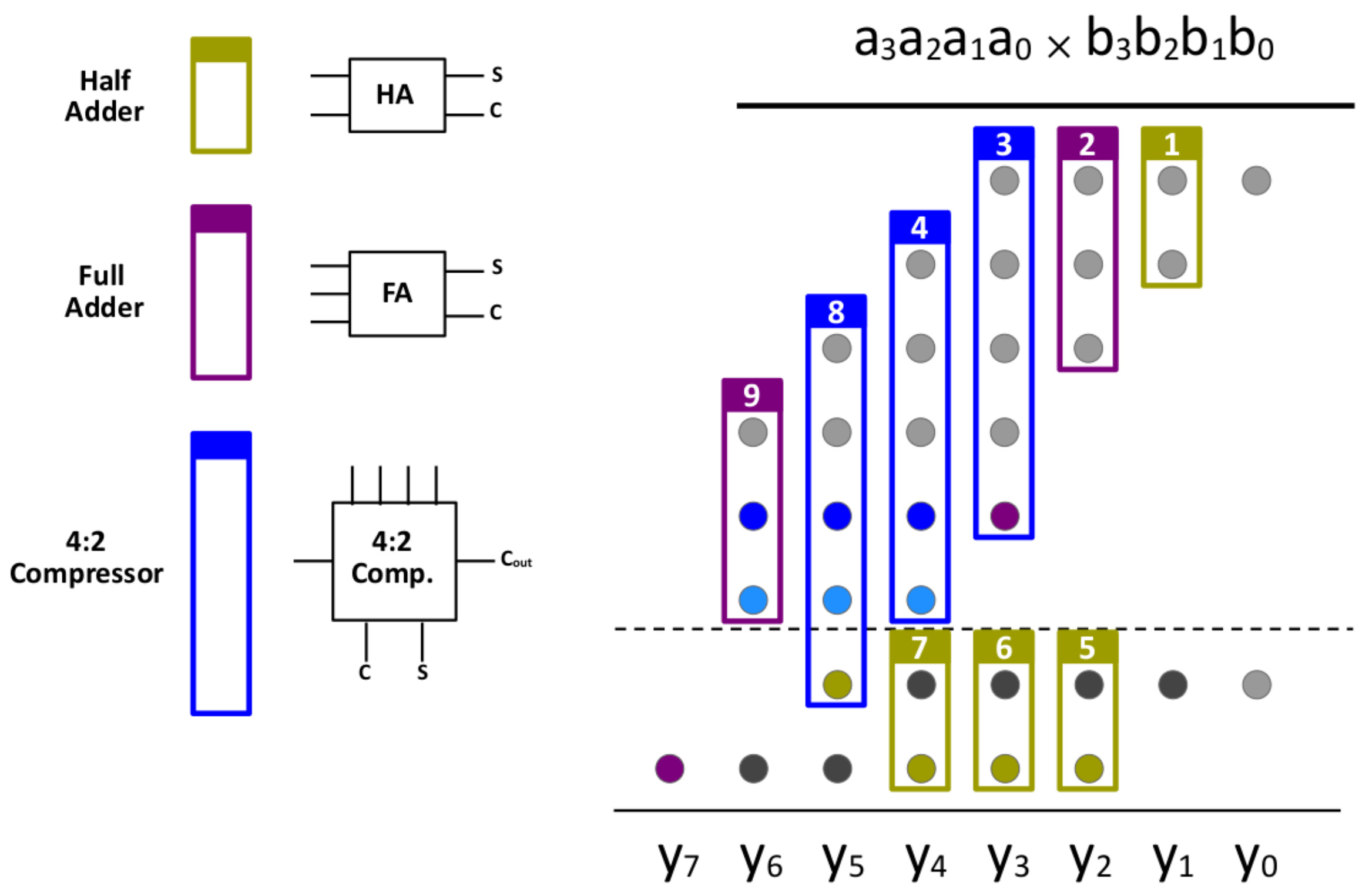}
	\caption{Proposed $4\times 4$ multiplier based on 4:2 compressors, full adders, and half adders.}
	\label{fig7}
\end{figure}
 
 The dot notation of Phase \Romannum{2} of the proposed 4-bit multiplier is depicted in Fig. \ref{fig7}. The 4-bit multiplier consists of four rows and seven columns of partial products. In the proposed multiplier, the carry bits resulting from the addition in column $k$ are added with the partial products of column $k+1$ until the final product is obtained. Phase \Romannum{2} of the proposed 4-bit multiplier is implemented by applying four half adders, two full adders, and three 4:2 compressors based on IMPLY logic with serial architecture. Arithmetic circuits are numbered in the order of implementation depicted in Fig. \ref{fig7}. For instance, consider the 4:2 compressor number 4 with its five inputs, including partial products $a_{3}b_{1}$, $a_{2}b_{2}$, $a_{1}b_{3}$, and the outputs $C_{out}$ and $Carry$ of the 4:2 compressor number 3. The operations of the corresponding 4:2 compressor are executed based on the proposed algorithm in Table \ref{tab3}. At the end of the computations, its $C_{out}$ and $Carry$ outputs are the inputs of the 4:2 compressor number 8. Then, half adder number 5, whose inputs are the $C_{out}$ output of the half adder number 1 and the $Sum$ output of full adder number 2, is computed. The order of implementation of the arithmetic circuits in Phase \Romannum{2} is such that at the end of the computation of each column, the input memristors $a_{j}$ or $b_{i}$ are reused to store the bits of the final product. Excluding $b_{0}$, the other 17 memristors of Phase \Romannum{1} can be reused to implement the operations of Phase \Romannum{2}. Hence, no more memristor is needed. In Phase \Romannum{1}, the memristors that stored the partial products are the input memristors of Phase \Romannum{2}. According to the considered algorithm, memristors $S_{1}$ and $b_{3}$ remain unused at the end of Phase \Romannum{1}. So, these two memristors are used as work memristors of half adder number 1. The $Sum$ output of the half adder number 1 equals the bit $y_{1}$ of the final product. At the end of the operations of this circuit, the $Sum$ output is stored in memristor $b_{3}$, and memristors $S_{1}$ and $S_{5}$ remain unused. As discussed earlier in previous sections, the full adder and proposed 4:2 compressor algorithms are written as the logic state of the outputs is saved in the input memristors. Hence, the memristors $S_{1}$ and $S_{5}$ can be used as work memristors of all 4:2 compressors and full adders. The memristor $S_{1}$ has also been used as the work memristor of all half adder circuits. The $Sum$ output of each half adder equals one bit of the final product. According to the half adder’s algorithm in Table \ref{tab5}, the $Sum$ output is stored in the work memristor. So, one of the input memristors of the multiplier is applied as the corresponding work memristor of the half adder. The half-adder circuit is implemented with 4 memristors and 12 computational steps. In addition, 5 memristors and 22 computational steps are required to implement the full adder circuit \cite{ref9}. Seven memristors and 44 steps are needed to implement the proposed 4:2 compressor. Finally, by overlapping operations and reusing the memristors of Phase \Romannum{1}, Phase \Romannum{2} of the proposed 4-bit multiplier designed based on the proposed 4:2 compressor has been implemented with 17 memristors and in 224 computational steps. Therefore, 18 memristors and 304 computational steps are needed to implement and execute the proposed 4-bit multiplier.

\begin{table}
  \begin{center}
    \caption{IMPLY-based half adder algorithm with serial implementation}
    \label{tab5}
    \resizebox{0.48\textwidth}{!}{
    \begin{tabular}{|c|c|c|}
      \hline
      \textbf{Step} & \textbf{Operation} & \textbf{Equivalent Logic} \\ \hline
			\textbf{1} & $S_{1}=0$ & FALSE($S_{1}$) \\ \hline
                \textbf{2} & $S_{2}=0$ & FALSE($S_{2}$) \\ \hline
                \textbf{3} & $(A \to 0) \equiv (A \to S_{1})=S_{1}^{'}$ &  $S_{1}^{'}=\overline{A}$\\ \hline
                \textbf{4} & $(B \to 0) \equiv (B \to S_{2})=S_{2}^{'}$ &  $S_{2}^{'}=\overline{B}$\\ \hline
                \textbf{5} & $(S_{1}^{'}\to S_{2}^{'})=S_{2}^{''}$ & $S_{2}^{''}=A+\overline{B}$ \\ \hline
                \textbf{6} & $(B\to S_{1}^{'})=S_{1}^{''}$ & $S_{1}^{''}=\overline{A}+\overline{B}$ \\ \hline
                \textbf{7} & $(A\to B)=B^{'}$ & $B^{'}=\overline{A}+B$ \\ \hline
                \textbf{8} & $A=0$ & FALSE($A$) \\ \hline
                \textbf{9} & $(S_{1}^{''}\to 0) \equiv (S_{1}^{''}\to A)=A^{'}$ & \cellcolor{tablecolor}$A^{'}=A \cdot B=\bm{\textcolor{red}{C_{out}}}$ \\ \hline
                \textbf{10} & $S_{1}=0$ & FALSE($S_{1}$) \\ \hline
                \textbf{11} & $(S_{2}^{''} \to 0)\equiv(S_{2}^{''}\to S_{1})=S_{1}^{'}$ & $S_{1}^{'}=\overline{A}\cdot B$  \\ \hline
                \textbf{12} & $(B^{'}\to S_{1}^{'})=S_{1}^{''}$ & \cellcolor{tablecolor}$S_{1}^{''}=A \cdot \overline{B}+\overline{A}\cdot B$ \\ 
                & & \cellcolor{tablecolor}$ = A \oplus B = \bm{\textcolor{red}{Sum}}$ \\ \hline
    \end{tabular}}
  \end{center}
\end{table}      

 \subsubsection{8$\times$8 Multiplier}\label{sec322}
 The 8-bit multiplier consists of two 8-bit inputs, $a_{7...0}$ as multiplicand and $b_{7...0}$ as multiplier. Hence, 16 input memristors are required to implement the proposed 8-bit multiplier. Phase \Romannum{1} of the proposed 8-bit multiplier comprises 64 AND gates implemented and executed serially. Phase \Romannum{1} of the proposed n-bit multiplier is implemented by applying the $n^2$ number of serial AND gates. Like the 4-bit multiplier, the Phase \Romannum{1} algorithm is implemented as the last partial product generated in the $i^{th}$ row $(i=0, 1, ..., 6)$, which is stored in the input memristor $b_{i}$. Also, all partial products of the last row are stored in the input memristors $a_{j}$ $(j=0, ..., 7)$. Thus, the partial product $a_{0}b_{0}$ (equal to bit $y_{0}$ of the final product) is stored in the memristor $b_{0}$, the partial products $a_{7}b_{i}$ $(i=1, 2, ..., 6)$ are each stored in the corresponding memristor $b_{i}$, and the partial products of the last row, $a_{j}b_{7}$ $(j=0, 1 , ..., 7)$, are each stored in the corresponding input memristor $a_{j}$. The logic state of other partial products is stored in the work memristors due to the dependency of the operations on the initial logic state of the input memristors. Hence, the number of required work memristors reaches 49. Also, one work memristor is shared between all AND gates to perform logic operations. Therefore, 66 memristors, including 16 input and 50 work memristors, are needed to implement Phase \Romannum{1} of the proposed 8-bit multiplier. Since each AND gate is computed in 5 steps, Phase \Romannum{1} of the proposed 8-bit multiplier executes in 320 computational steps. The proposed design generally implements Phase \Romannum{1} of the n-bit multiplier with $n^2+2$ number of memristors and $5n^2$ computational steps.
 
 \begin{figure}[h]
	\centering
	\includegraphics[scale=0.8]{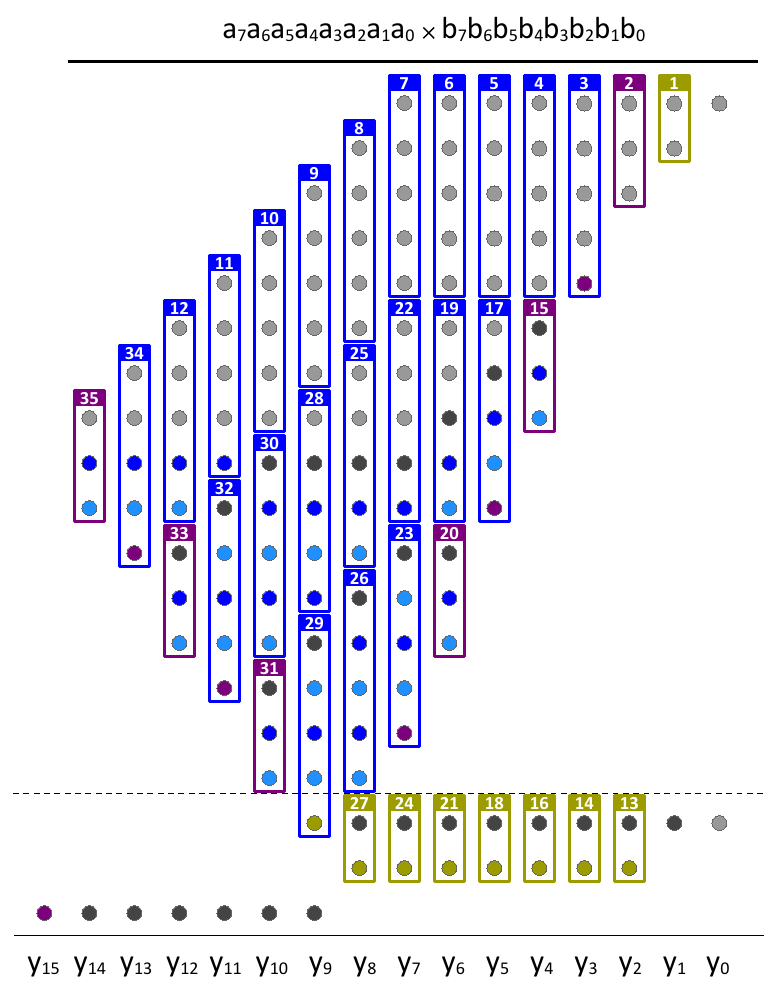}
	\caption{Proposed $8 \times 8$ multiplier based on 4:2 compressors, full adders, and half adders.}
	\label{fig8}
\end{figure}

 As presented in Fig. \ref{fig8}, in Phase \Romannum{2} of the proposed 8-bit multiplier, the PPR and addition operations are implemented by applying eight half adders, six full adders, and twenty-one 4:2 compressors until the final product, $y_{15...0}$, is attained. Thus, $n$ half adders, $n-2$ full adders, and $\frac{1}{2}(n^2-3n+2)$ 4:2 compressors are required to implement Phase \Romannum{2} of the proposed n-bit multiplier. The number of 4:2 compressors applied in the proposed n-bit multiplier is computed based on the ``\textit{Gauss sum}" method and the placement of the proposed 4:2 compressors in phase \Romannum{2} of the multipliers. Each arithmetic circuit is numbered in the order of implementation as shown in Fig. \ref{fig8}. In Phase \Romannum{2} of the proposed algorithm, the $Sum$ output of the last arithmetic block (4:2 compressor, full adder, or half adder) of each column is stored in one of the input memristors of the multiplier. At the end of Phase \Romannum{1} of multiplication, memristors $b_{7}$ and $S_{1}$ remain unused, which are applied as work memristors of half adder number 1. So, the $Sum$ output of this circuit, equivalent to bit $y_{0}$, can be stored in input memristor $b_{7}$, and the memristor $S_{1}$ is shared between other arithmetic blocks of the multiplier as a work memristor. Moreover, according to Phase \Romannum{1} of the proposed algorithm, the half adder number 1 inputs, i.e., partial products $a_{1}b_{0}$ and $a_{0}b_{1}$, are saved in memristors $S_{2}$ and $S_{9}$, respectively. At the end of the computation of this half adder, the $C_{out}$ output is stored in the memristor $S_{2}$, and the memristor $S_{9}$ remains unused. Therefore, the memristor $S_{9}$ can be used as a work memristor for all 4:2 compressors and full adders. The number of memristors and computational steps needed to implement each arithmetic circuit based on serial IMPLY logic were investigated earlier. Finally, 65 memristors and 1152 computational steps are needed to implement Phase \Romannum{2} of the proposed 8-bit multiplier (considering the reuse of Phase \Romannum{1} memristors). The execution of Phase \Romannum{2} of the proposed n-bit multiplier takes $22n^2-32n$ computational steps. 
 
 The number of the computational steps of the proposed n-bit multiplier equals the sum of the computational steps of all logic gates and arithmetic blocks, including AND gates, half adders, full adders, and 4:2 compressors, as given here in Eq. \ref{eq15n}. \textit{CS} refers to the number of computational steps in Eq. \ref{eq15n}.
 
 \begin{multline}
    CS_{multiplier}=(\# AND\times CS_{AND})+(\# HA\times CS_{HA})+(\# FA\times CS_{FA})+(\# 4:2 Comp\times CS_{4:2 Comp})
    \label{eq15n}     
 \end{multline}

\section{Simulation and Evaluation of The Proposed Designs} \label{sec4}
\subsection{Simulation Setup} \label{sec41}
 The proposed 4:2 compressor is simulated using SPICE to evaluate the circuit’s functionality. The Voltage-controlled ThrEshold Adaptive Memristor (VTEAM) model was conducted to simulate the proposed circuit \cite{ref41}. The VTEAM model builds upon the threshold voltage, $V_{th}$. The memristor resistance changes if the voltage across the memristor exceeds its threshold value. Otherwise, the memristor does not switch, and its resistance remains unaltered \cite{ref41}. The values of VTEAM model parameters are itemized in Table \ref{tab6} \cite{ref28}. Also, the initial parameters of the circuit, including the applied voltage of the memristors, are adjusted based on IMPLY logic setup conditions \cite{ref28, ref10}, which are:
 $\{V_{SET}=1V,~V_{RESET}=1V,~V_{COND}=900mV,~R_{G}=40K\Omega,~T_{pulse}=30\mu s\}$

\begin{table}
	\centering
	\caption{VTEAM model setup parameters \cite{ref28, ref10}}
	\scalebox{0.89}{
		\begin{tabular}{|c|c|c|c|c|c|c|} \hline 
			\textbf{Parameter} & $\bm{v_{on}}$ & $\bm{v_{off}}$ & $\bm{\alpha_{on}}$ & $\bm{\alpha_{off}}$ & $\bm{R_{on}}$ & $\bm{R_{off}}$    \\ \hline  
                \textbf{Value} & -10 mV & 0.7 V & 3 & 3 & 10 K$\Omega$ & 1 M$\Omega$ 
                \\ \hline 
                $\bm{K_{on}}$ & $\bm{K_{off}}$ & $\bm{\omega_{on}}$ & $\bm{\omega_{off}}$ & $\bm{\omega_{c}}$ & $\bm{a_{on}}$ & $\bm{a_{off}}$
                \\ \hline  
                -0.5 $\frac{nm}{s}$ & 1 $\frac{cm}{s}$ & 3 nm & 0 nm & 107 pm & 0 nm & 3 nm
                \\ \hline 
	\end{tabular}}
	\label{tab6}
\end{table}

 \subsection{Simulation Results} \label{sec42}
\begin{figure}[h]
 	\centering
 	\includegraphics[scale=0.2]{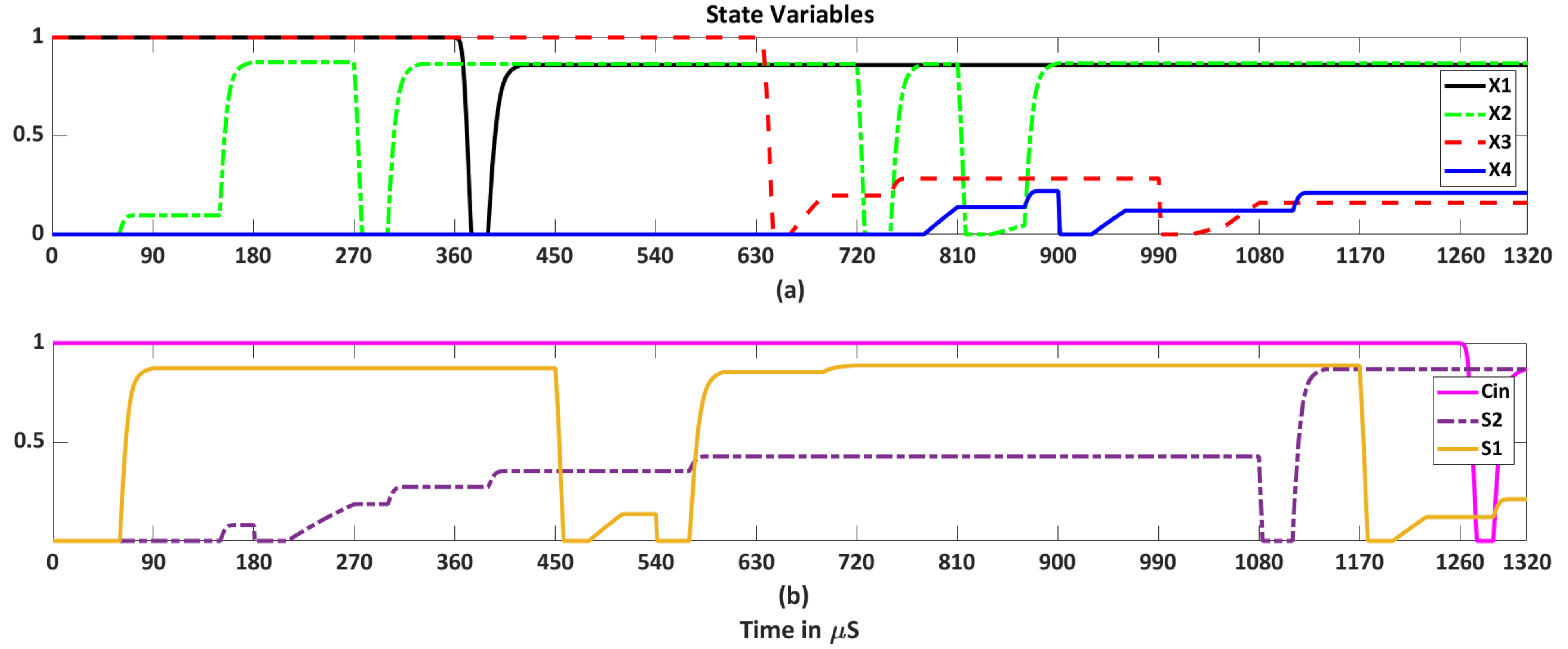}
 	\caption{Simulation result of 4:2 compressor for input state = ``11010'' with memristors: (a) $X_{1}$, $X_{2}$, $X_{3}$, $X_{4}$, and (b) $C_{in}$, $S_{1}$, $S_{2}$.}
 	\label{fig9}
 \end{figure}

\begin{figure}[h]
 	\centering
 	\includegraphics[scale=0.2]{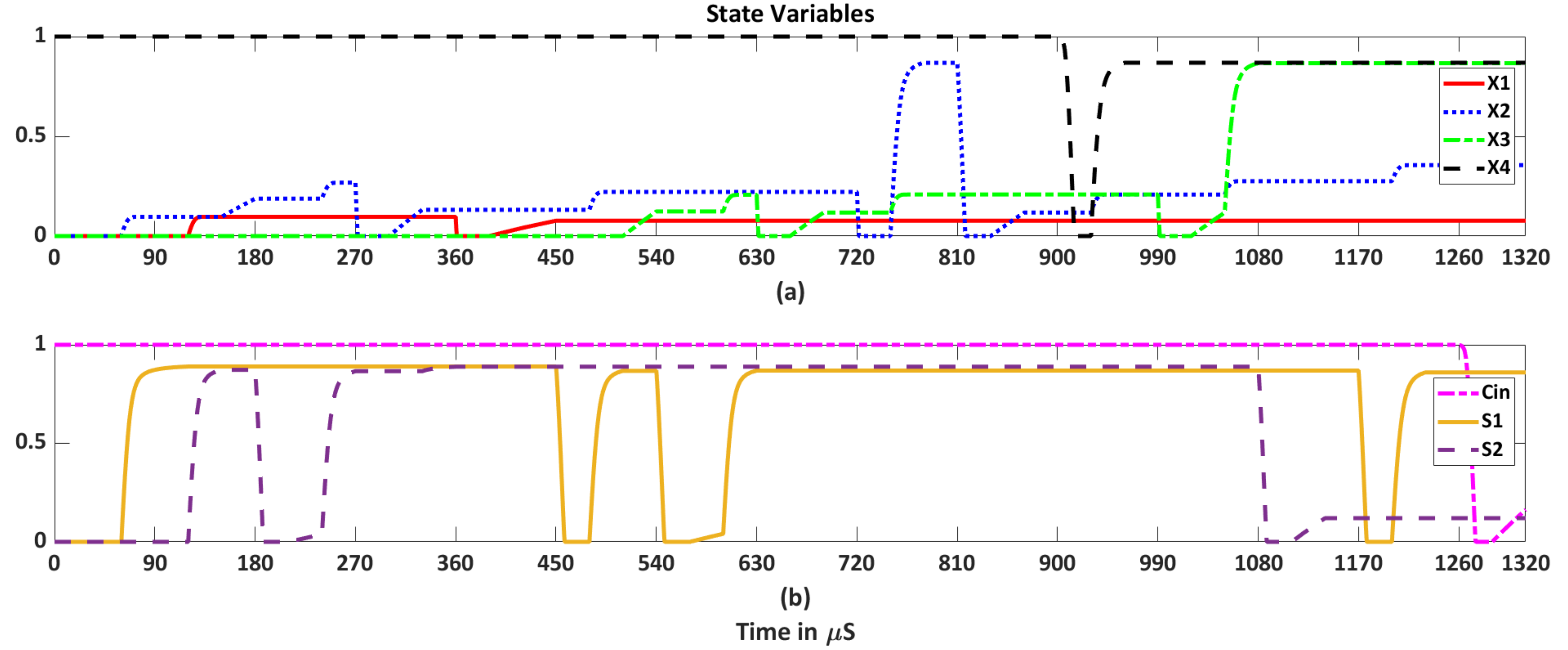}
 	\caption{Simulation result of 4:2 compressor for input state = ``10001'' with memristors: (a) $X_{1}$, $X_{2}$, $X_{3}$, $X_{4}$, and (b) $C_{in}$, $S_{1}$, $S_{2}$.}
 	\label{fig10}
 \end{figure}

\begin{figure}[h!]
 	\centering
 	\includegraphics[scale=0.2]{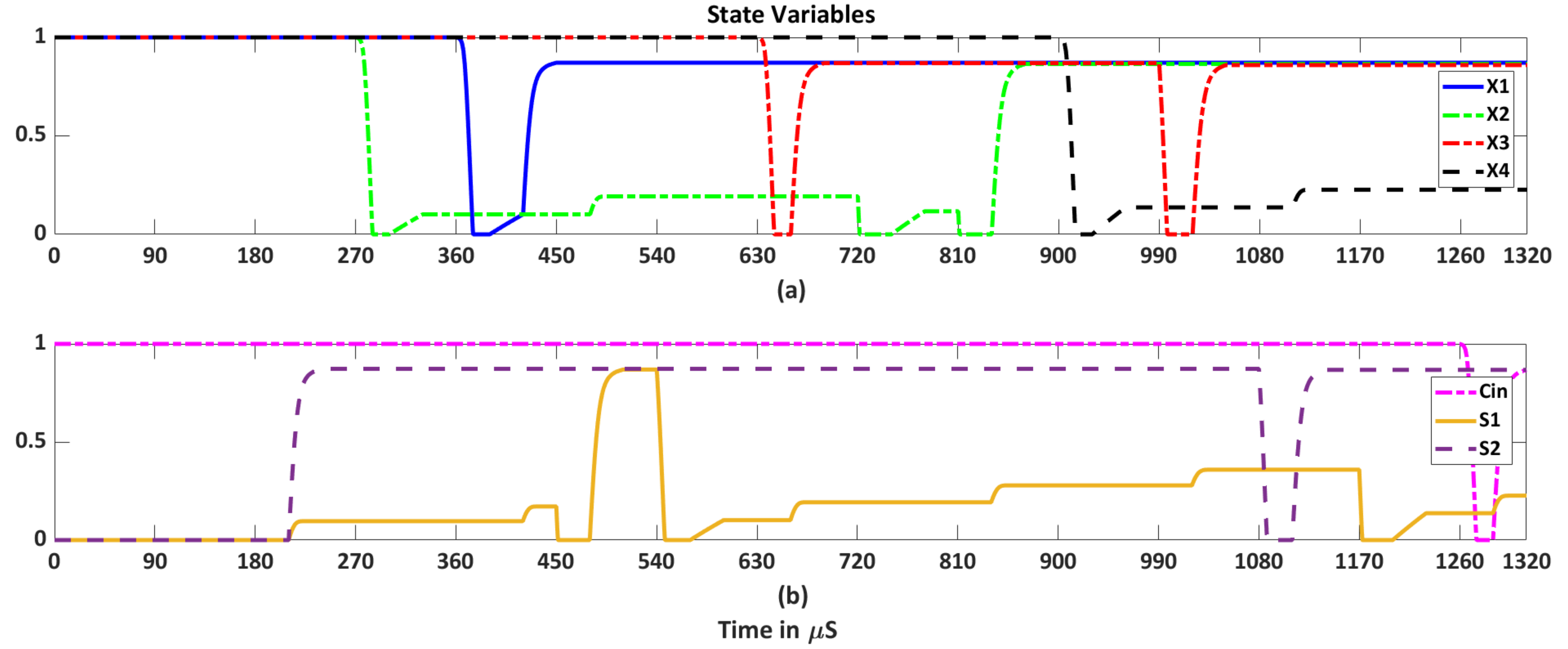}
 	\caption{Simulation result of 4:2 compressor for input state = ``11111'' with memristors: (a) $X_{1}$, $X_{2}$, $X_{3}$, $X_{4}$, and (b) $C_{in}$, $S_{1}$, $S_{2}$.}
 	\label{fig11}
 \end{figure}

 The proposed 4:2 compressor and its implementation algorithm (see Table \ref{tab3}) underwent simulation for all 32 input states to evaluate the circuit functionality. The proposed algorithm requires 44 computational steps, each taking 30 $\mu s$ ($T_{pulse}$). The proposed 4:2 compressor is simulated in the time interval (0 $\mu s$–1320 $\mu s$). Fig. \ref{fig9}, Fig. \ref{fig10}, and Fig. \ref{fig11} present the waveform of memristors resistance for input states “11010”, “10001”, and “11111”, respectively, for instance. The proposed 4:2 compressor computes the $C_{out}$ output in 450 $\mu s$ and stores it in the memristor $X_{1}$. Moreover, the $Carry$ output, saved in the memristor $X_{3}$, and the $SUM$ output, saved in the memristor $C_{in}$, are computed in 1080 $\mu s$ and 1320 $\mu s$, respectively. Furthermore, the energy consumption of the proposed 4:2 compressor circuit is computed. First, each memristor’s energy consumption is measured at the end of each input state’s simulation using SPICE based on (\ref{eq16}). Then, obtained energy values are summed for each input state. Finally, the total energy consumption of the proposed 4:2 compressor is obtained from the average energy consumption of all input states \cite{ref10,ref42}, which is 3.76 nJ.
 
 \begin{equation}
     E=\int P_{avg}d_{t}=P_{avg}\cdot t
     \label{eq16}
 \end{equation}
 To estimate the energy consumption of the proposed multiplier, first, for each arithmetic block, the proposed 4:2 compressor, full adder, and half adder, as well as the AND gate, are simulated using SPICE. The energy consumption of each arithmetic element is obtained as given in Table \ref{tab7}. Then, the total energy consumption of the multiplier is estimated according to (\ref{eq17}). The total energy consumption of the proposed 4, 8, and 16-bit multipliers is 24.35 nJ, 119 nJ, and 521 nJ, respectively.
 \begin{equation}
     E_{multiplier}=(\# AND\times E_{AND})+(\# HA\times E_{HA})+(\# FA\times E_{FA})+(\# 4:2 Comp\times E_{4:2 Comp})
     \label{eq17}     
 \end{equation}
 
 \begin{table}
 	\centering
 	\caption{Energy consumption results for computing elements}
 	\scalebox{0.89}{
 		\begin{tabular}{|c|c|c|c|}
 			\hline
 			\multicolumn{4}{|c|}{\textbf{Energy Consumption (nJ)}} \\ \hline
 			\textbf{AND gate} & \textbf{Half adder} & \textbf{Full adder} & \textbf{Proposed 4:2 compressor} \\ \hline
 			0.33 & 1.02 & 1.85 & 3.76 \\
 			\hline
 	\end{tabular}}
 	\label{tab7}
 \end{table}
 
\begin{table}
  \begin{center}
    \caption{The comparison of IMPLY-based 4:2 compressor designs}
    \label{tab8}
    \resizebox{0.7\textwidth}{!}{
    \begin{tabular}{|c|c|c|c|c|c|c|}
      \hline
        \multirow{3}{*}{\textbf{Design}} & \multicolumn{2}{|c|}{\textbf{Number of}} & \multicolumn{2}{|c|}{\textbf{Number of}} & \multicolumn{2}{|c|}{\textbf{Energy}} \\
        & \multicolumn{2}{|c|}{\textbf{Memristors}} & \multicolumn{2}{|c|}{\textbf{Steps}} & \multicolumn{2}{|c|}{\textbf{Consumption}}\\ \cline{2-7}
        & \textbf{Total} & \textbf{Imp.} & \textbf{Total} & \textbf{Imp.} & \textbf{Total} & \textbf{Imp.} \\ \hline
        \textbf{Proposed NAND based} & 7 & - & 44 & - & 3.76 nJ & - \\ \hline
        \textbf{Serial XOR/MUX \cite{ref23}} & 7 & 0 & 52 & 15\% & 4.54 nJ & 17\% \\ \hline
        \textbf{Parallel XOR/MUX \cite{ref23}} & 11 & 36\% & 26 & -41\% & $NA^*$ & $NA^*$ \\ \hline             
    \end{tabular}}
  \end{center}
     {\raggedright $^*$The source does not provide expressions for the energy consumption. \par}
\end{table}

\begin{table}
	\begin{center}
		\caption{The comparison of IMPLY-based multipliers for $n=4$ and $n=8$}
		\label{tab9}
		\resizebox{1\textwidth}{!}{
			\begin{tabular}{|c|c|c|c|c|c|c|c|c|c|c|c|c|c|c|c|}
				\hline
				\multirow{2}{*}{\textbf{Design}} & \multicolumn{5}{c|}{\textbf{Number of Memristors}} & \multicolumn{5}{c|}{\textbf{Number of steps}} & \multicolumn{5}{c|}{\textbf{Energy Consumption}} \\ \cline{2-16}
				& \textbf{Total} & \textbf{n=4} & \textbf{Imp.} & \textbf{n=8} & \textbf{Imp.} & \textbf{Total} & \textbf{n=4} & \textbf{Imp.} & \textbf{n=8} & \textbf{Imp.} & \textbf{Total} & \textbf{n=4} & \textbf{Imp.} & \textbf{n=8} & \textbf{Imp.} \\ \hline
				\textbf{Proposed} & $n^{2}+2$ & 18 & - & 66 & - & $27n^{2}-32n$ & 304 & - & 1472 & - & $2.21n^{2}-2.8n-0.05$ & 24.35 & - & 119 & - \\ \hline
				\textbf{based on} & \multirow{2}{*}{$n^{2}+2$} & \multirow{2}{*}{18} & \multirow{2}{*}{0} & \multirow{2}{*}{66} & \multirow{2}{*}{0} & \multirow{2}{*}{$31n^{2}-44n+8$} & \multirow{2}{*}{328} & \multirow{2}{*}{7.3\%} & \multirow{2}{*}{1640} & \multirow{2}{*}{10\%} & \multirow{2}{*}{$2.6n^{2}-3.94n+0.84$} & \multirow{2}{*}{26.68} & \multirow{2}{*}{8.7\%} & \multirow{2}{*}{135.7} & \multirow{2}{*}{12\%} \\ 
				\textbf{4:2 compressor \cite{ref23}} & &  &  &  &  &  &  &  &  &  &  &  &  &  &  \\ \hline
		\end{tabular}}
	\end{center}
\end{table}

 \subsection{Evaluation and Comparison} \label{sec43}
 The efficiency of the arithmetic circuit is evaluated based on essential criteria such as energy consumption, area, and speed (computational delay). The number of memristors and computational steps of the memristive IMPLY-based circuit are examined to evaluate the area and delay, respectively. The proposed 4:2 compressor is evaluated regarding the energy consumption, the number of memristors, and computational steps and compared with previous IMPLY-based 4:2 compressors, including the XOR/MUX design \cite{ref23}. Table \ref{tab8} presents the evaluation results of the proposed 4:2 compressor compared to SOA. The proposed 4:2 compressor improves 36\% over the parallel XOR/MUX design \cite{ref23} regarding the number of required memristors. Also, it is improved by 15\% compared to the serial XOR/MUX design \cite{ref23} in terms of computational steps. The proposed 4:2 compressor circuit and serial XOR/MUX design are implemented with 7 memristors, the minimum number of memristors required to implement the intended serial design. The energy consumption of the 4:2 compressor using the XOR/MUX design is not reported in \cite{ref23}. Therefore, the serial XOR/MUX design of the 4:2 compressor \cite{ref23} was simulated under the same initial conditions as the proposed circuit, and its energy consumption was calculated. The energy consumption of the proposed 4:2 compressor is 17\% less than the serial XOR/MUX design \cite{ref23}. This improvement in energy consumption is due to the fewer computational steps of the proposed circuit, hence, less switching of memristors. As discussed in Section \ref{sec1}, the memristor $q$ is applied as an input and output memristor in the IMPLY logic operation. The switching of memristor $q$ significantly affects the circuit's energy consumption. The memristor $q$ switches when the resistance state variable of the memristor changes from $R_{OFF}$ to $R_{ON}$ when $p$ and $q$ are ‘0’.

 %There are few existing works to evaluate and compare with the proposed 4:2 compressor; the only memristive IMPLY-based 4:2 compressor is presented in \cite{ref23}. As mentioned in Section \ref{sec2}, one of the 4:2 compressor designs is based on the serial connection of two full adders. Therefore, in addition to the XOR/MUX design \cite{ref23}, the fast serial full adder with 22 computational steps \cite{ref9} was investigated. The 4:2 compressor circuit is implemented containing the corresponding full adder with 7 memristors and in 44 computational steps. The proposed NAND-based 4:2 compressor and the presented full adder-based design require the same number of memristors and computational steps to be implemented. The energy consumption of the proposed NAND-based 4:2 compressor is improved by 1\% when compared with the full adder-based design.

Besides the proposed 4:2 compressor, the proposed 4-bit and 8-bit multipliers were evaluated regarding the number of memristors, computational steps, and energy consumption. No existing work presents serial crossbar array compatible multipliers to compare with the proposed multipliers. Therefore, the multipliers are implemented containing the 4:2 compressor based on XOR/MUX design \cite{ref23} in serial architecture (based on structures in Fig. \ref{fig7} and Fig. \ref{fig8}). The computational steps and energy consumption of these multipliers are calculated by the same method applied for the proposed 4-bit and 8-bit multipliers according to (\ref{eq15n}) and (\ref{eq17}). The evaluation results for 4, 8, and n-bit multipliers are presented in Table \ref{tab9}. The proposed 4-bit and 8-bit multipliers are improved by 7.3\% and 10\%, respectively, in terms of the computational steps, compared to the serial 4:2 compressor \cite{ref23} based multipliers. Moreover, the proposed 4-bit and 8-bit multipliers reduce energy consumption by 8.7\% and 12\%, respectively, compared to the mentioned 4:2 compressor \cite{ref23} based multipliers.

 \section{Conclusion} \label{sec5}
 A memristive IMPLY-based 4:2 compressor circuit was implemented with serial architecture for IMC. This arithmetic circuit is an important cell of the multiplier’s PPR stage for modern processing applications. The proposed 4:2 compressor was applied to implement new 4-bit and 8-bit multipliers. Besides the proposed 4:2 compressor, these multipliers were analyzed and evaluated regarding the number of required memristors, computational steps, and energy consumption. The memristors of the proposed 4:2 compressor circuit are reduced by 36\% compared to the parallel XOR/MUX design. Also, compared to the serial XOR/MUX design, it has been improved by 17\% and 15\% regarding the energy consumption and computational steps, respectively. The number of computational steps in the proposed 4-bit and 8-bit multipliers has been reduced by 7.3\% and 10\%, respectively, compared to the multipliers designed with serial memristive XOR/MUX-based 4:2 compressor. Furthermore, compared to the SOA, the energy consumption of the proposed 4-bit and 8-bit multipliers has been improved by 8.7\% and 12\%, respectively.
 
\subsection*{Author Contributions}
\noindent \textbf{Bahareh Bagheralmoosavi}: Conceptualization, Methodology, Software, Validation, Formal Analysis, Investigation, Data Curation, Writing - Original Draft, Writing - Review \& Editing, Visualization.

\noindent \textbf{Seyed Erfan Fatemieh}: Conceptualization, Methodology, Software, Validation, Formal Analysis, Investigation, Data Curation, Writing - Original Draft, Writing - Review \& Editing, Visualization. 

\noindent \textbf{Mohammad Reza Reshadinezhad}: Software, Validation, Formal Analysis, Investigation, Resources, Writing - Review \& Editing, Supervision, Project Administration. 

\noindent \textbf{Antonio Rubio}: Validation, Formal Analysis, Writing - Review \& Editing, Funding acquisition.

\subsection*{Data Availability Statement}
Data is contained within the article.

\section*{Declarations}
\subsection*{Conflict of interest}
The authors declare no conflicts of interest.

\section*{Acknowledgements}
\noindent This work was supported in part by the Spanish Ministry of Science, Innovation and Universities and through ERDF/EU funds under Grant PID2022-141391OB-C22/ AEI/10.13039/501100011033.

\bibliographystyle{ieeetr} 
\bibliography{Manuscript}

\section*{Biography}
\noindent \textbf{\\ Bahareh Bagheralmoosavi} was born in Karaj, Iran, in 1996. She received her B.Sc. degree in Electrical Engineering from the University of Mohaghegh Ardabili, Ardabil, in 2019, and her M.Sc. degree in computer architecture from the University of Isfahan, Isfahan, in 2023. Her research interests include computer arithmetic, digital VLSI, emerging technology concerning memristors, and in-memory computing. \par

\noindent \textbf{\\ Seyed Erfan Fatemieh} was born in Isfahan, Iran, in 1996. He received his B.Sc. degree in Computer Engineering and M.Sc. degree in Computer Architecture from the University of Isfahan, Isfahan, Iran, in 2018 and 2020. He is currently a Ph.D. candidate in Computer Architecture at the University of Isfahan, Isfahan, Iran. His research interests include In-memory Computing, Digital VLSI, Computer Arithmetic, and Quantum Computing and Reversible Circuits.\par

\noindent \textbf{\\ Mohammad Reza Reshadinezhad} was born in Isfahan, Iran, in 1959. He received his B.S. and M.S. degrees from the Electrical Engineering Department of the University of Wisconsin, Milwaukee, USA, in 1982 and 1985, respectively. He has been in the position of lecturer as faculty of computer engineering at the University of Isfahan since 1991. He also received a Ph.D. Degree in computer architecture from Shahid Beheshti University, Tehran, Iran, in 2012. He is currently an Associate Professor in the Faculty of Computer Engineering at the University of Isfahan. His research interests are Digital Arithmetic, Nanotechnology concerning CNTFET, VLSI Implementation, and Cryptography.\par

\noindent \textbf{\\ Antonio Rubio} received the M.Sc. and Ph.D. degrees from the Industrial Engineering Faculty, Polytechnic University of Catalonia (UPC), Barcelona, Spain. He has been an Associate Professor with the Electronic Engineering Department, UPC, and a Professor with the Physics Department, Balearic Islands University, Palma, Spain. He is currently a Full Professor in electronic technology with the Telecommunication Engineering Faculty, UPC. His research interests include VLSI design and test, device and circuit modeling, high-speed circuit design, and new emerging nanodevices and nanoarchitectures. He is an Associate Editor of the IEEE Transactions on Computers, a Senior Editor of the IEEE Transactions of Nanotechnology, and the Chair of IEEE Nano Giga TC. He was the IEEE Computer Society Integrity Chair, from 2019 to 2021.

\end{document}